\documentclass[nofootinbib,aps,showpacs,superscriptaddress,preprintnumbers,epsf,psf]{revtex4-2}
\usepackage[dvips,final]{graphicx}
\usepackage[english]{babel}[2020/02/03]
\usepackage{pstricks}
\usepackage{epsf}
\usepackage{amsmath}
\usepackage{amsfonts}
\usepackage{amssymb}
\usepackage{dcolumn}% Align table columns on decimal point
\usepackage{bm}% bold math
\usepackage{mathrsfs}
\usepackage{slashed}
\newcommand\underrel[2]{\mathrel{\mathop{#2}\limits_{#1}}}

\usepackage[colorlinks = true,
linkcolor = blue,
urlcolor  = blue,
citecolor = blue,
anchorcolor = blue]{hyperref}
\usepackage{physics}

\begin{document}
	
%\begin{flushright}
%\vskip1cm
%\end{flushright}
	
\title{Viscosity, entanglement and acceleration  \vspace{1.0 cm}}
%\author{}
%\email{}
%\affiliation{\vspace{0.8 cm}}

\author{D. D. Lapygin}
\email{dmitrijlapygin@gmail.com}
\affiliation{Southern Federal University, Zorge  5,  Rostov-on-Don, 344090, Russia}

\author{G. Yu. Prokhorov}
\email{prokhorov@theor.jinr.ru}
\affiliation{Joint Institute for Nuclear Research, Joliot-Curie 6, Dubna, 141980, Russia}
\affiliation{NRC Kurchatov Institute, Moscow, Russia}

\author{O. V. Teryaev}
\email{teryaev@jinr.ru}
\affiliation{Joint Institute for Nuclear Research, Joliot-Curie 6, Dubna, 141980, Russia}
\affiliation{NRC Kurchatov Institute, Moscow, Russia}

\author{V. I. Zakharov}
\email{vzakharov@itep.ru}
\affiliation{NRC Kurchatov Institute, Moscow, Russia}
\affiliation{Joint Institute for Nuclear Research, Joliot-Curie 6, Dubna, 141980, Russia}
%\affiliation{Pacific Quantum Center,
%Far Eastern Federal University, 10 Ajax Bay, Russky Island, Vladivostok 690950, Russia}

\begin{abstract}\vspace{0.2 cm}
The Minkowski vacuum in an accelerated frame behaves like a fluid that has not only a finite temperature due to the Unruh effect, but also a finite shear viscosity. Moreover, the ratio of this viscosity to the entropy density exactly satisfies the Kovtun-Son-Starinets (KSS) bound, inspired by the string theory $ \eta/s=1/4\pi $. The origin of this viscosity is purely kinematical and is believed to be related to entanglement introduced by the Rindler horizon. We directly calculate the viscosity, entropy density, and their ratio for massless fields with spins 1/2 and 1. We show that locally the ratio of viscosity to entropy density can be below the limiting value $ 1/4\pi $ at distances of the order of the thickness of the membrane corresponding to the stretched horizon, and is described by the universal function for different spins. In particular, on the membrane surface $ \eta/s=1/8\pi $.
% It is also concluded that the limitation on the minimum temperature in an accelerated system apparently coincides with the limitation on the minimum (shear and bulk) viscosity.
\end{abstract}

\maketitle
\tableofcontents
%================
\section{Introduction}
\label{sec intro}
%================

The discovery of the quark-gluon plasma at the RHIC accelerator \cite{STAR:2005gfr} as a new state of matter, stimulated the study of novel phenomena arising at the border of gravity with thermodynamics and hydrodynamics.
%, see e.g. the review \cite{Schafer:2009dj}
An intriguing direction is the search for the effects in the medium associated with black holes. This relationship between such different areas can be substantiated either using holographical dualities \cite{Casalderrey-Solana:2011dxg}, or within the framework of scenarios of emergent gravity, see for example \cite{Verlinde:2010hp, Jacobson:1995ab, Prokhorov:2022udo, Prokhorov:2019yft}.

One of the foremost manifestations of the physics of black holes in the medium is the universal lower bound on the ratio of shear viscosity $ \eta $ to entropy density $ s $
\begin{eqnarray} 
\frac{\eta}{s} \geqslant \frac{1}{4 \pi} \,.
\label{nsmain}
\end{eqnarray}
A similar constraint on bulk viscosity $ \zeta $ was also predicted \cite{Buchel:2007mf}
\begin{eqnarray}
\frac{\zeta}{\eta} \geqslant 2 \left(\frac{1}{p}-c_s^2\right) \,,
\label{zs}
\end{eqnarray}
where $ p $ is the number of space dimensions (in the usual four-dimensional Minkowski space $ p=3 $), and $ c_s $ is the speed of sound. 

Initially, the relation for viscosity was obtained for the black hole membrane (see \cite{Parikh:1997ma} and references therein), the viscosity of which is at the lower limit of (\ref{nsmain})
\begin{eqnarray} 
\frac{\eta}{s} = \frac{1}{4 \pi} \,.
\label{main}
\end{eqnarray}

After that, using methods of string theory \cite{Kovtun:2004de, Kovtun:2003wp, Policastro:2001yc, Policastro:2002se} and the holographic principle, the existence of a universal constraint (\ref{nsmain}) was put forward. The bound (\ref{nsmain}) is often called the KSS bound (by the names of the authors of \cite{Kovtun:2004de}, P. K. Kovtun, D. T. Son and A. O. Starinets). It is believed that quark-gluon plasma is close to this bound \cite{Nakamura:2004sy, Teaney:2009qa, Schafer:2009dj, Arslandok:2023utm}. To date, the ratio $ \eta/s $ has been studied very intensively from very different angles and in different systems, including cold atoms \cite{Schafer:2007pr}, lattice QCD \cite{Nakamura:2004sy, Astrakhantsev:2017nrs}, nuclear matter \cite{Mondal:2017flr} and many other systems.
Note that in a number of cases there are indications of a possible violation of constraint (\ref{nsmain}), in particular, for the Gauss-Bonnet gravity \cite{Kats:2007mq, Brigante:2007nu, Cremonini:2011iq, Bravo-Gaete:2022lno} and in a strongly coupled anisotropic plasma \cite{Rebhan:2011vd}.	

In recent years, much attention has been focused on another class of phenomena that also demonstrate the relationship between hydrodynamics and fundamental physics. Experiments with heavy ions indicate that matter in their collisions is formed in a state with extremely high vorticity and acceleration \cite{STAR:2017ckg}. Due to this a class of novel, believed to be non-dissipative, transport phenomena, arise \cite{Zakharov:2012vv, Chernodub:2021nff, Kharzeev:2013jha}. For example, in a vortical medium, the Chiral Vortical Effect (CVE) was predicted, according to which there is a current along the vorticity. Moreover, it was shown that this effect is closely related to the chiral quantum anomaly \cite{Son:2009tf, Zakharov:2012vv}. Later, similar anomalous transport phenomena caused by the gravitational chiral anomaly were also predicted \cite{Prokhorov:2022udo, Landsteiner:2011cp}.

Various quantum statistical methods have been developed to study these effects of non-inertiality \cite{Buzzegoli:2017cqy, Gao:2012ix, Yang:2024sfp}. In particular, the famous Unruh effect \cite{Unruh:1976db} was demonstrated using a quantum-statistical approach \cite{Becattini:2017ljh, Prokhorov:2019yft}.
However, within this direction, the main efforts were aimed at studying equilibrium transport phenomena. And only recently have works appeared that study the effects associated with dissipation or the shear tensor \cite{Palermo:2024tza, Florkowski:2021xvy, Bhadury:2020cop}.

At this stage, a natural problem arises to also study dissipative phenomena in non-inertial systems. Interesting result in this direction was obtained in \cite{Chirco:2010xx}. It was shown that the thermal bath, considered in the Unruh effect, has not only a temperature, but also a finite viscosity. This viscosity appears from the leading divergent contribution coming from free fields and is assumed to be related to quantum entanglement. Moreover, for a Minkowski vacuum it satisfies the lower bound (\ref{main}). Note that the case of Rindler space, generally speaking, is not covered by the prediction \cite{Kovtun:2004de}, since, to the best of our knowledge, there is no holographic duality for field theory in Rindler space.

As already noted, the minimal viscosity in (\ref{main}) was first obtained using the membrane paradigm. In this context, it is assumed that the black hole horizon has a finite thickness $ l_c $, that is, we are talking about a stretched horizon. It can be shown that the membrane satisfies the Navier-Stokes equation and has shear viscosity (\ref{main}). Note however that there is a significant difference between the derivation \cite{Parikh:1997ma} and \cite{Chirco:2010xx}. The first one considers the (mostly ``classical'') properties of the membrane itself. In contrast, the derivation \cite{Chirco:2010xx} refers to a ``quantum'' fluid living above this membrane and is based on a direct calculation using the corresponding quantum theory of a (scalar) field.

In our current work, we continue to study the dissipative properties of different quantum fluids living above the Rindler membrane. We will generalize the derivation of \cite{Chirco:2010xx} to the case of massless fields with spins 1/2 and 1 and show that in all cases considered, the viscosity in the accelerated frame in the Minkowski vacuum state satisfies the lower bound (\ref{main}). We will also analyze the behavior of local quantities characterizing dissipative properties at a given distance from the Rindler horizon and show that (\ref{main}) is satisfied only on average, but is violated near the membrane. Namely, at the outer boundary of the membrane (that is, on the stretched horizon) the ratio of viscosity to entropy density turns out to be half the KSS bound $ \eta/s=1/8\pi $. In parallel, we will show how entanglement entropy can be calculated in all considered cases, using the approach to entropy from the relativistic spin hydrodynamics  \cite{Becattini:2023ouz, Becattini:2019poj}.

The paper has the following structure. First, in Section \ref{sec_kubo} we will talk about the methods used and the theories within which we work: Kubo formulas, the membrane paradigm, the approach to calculating entropy in Rindler space. Then, in Section \ref{sec_scalar}, we briefly review existing viscosity results for scalar fields. Section \ref{sec_dirac} presents original results for Dirac fields. Shear viscosity (and entropy) for electromagnetic fields are calculated in Section \ref{sec_electro}. Section \ref{sec_bulk} shows that in all cases considered, the bulk viscosity is zero. Section \ref{sec_disc} analyzes the results obtained, in particular, the origin of the viscosity for free fields in the Rindler space and violation of the ratio (\ref{main}) between local viscosity and entropy on the membrane surface. Section \ref{sec_concl} provides a conclusion and discusses possible further development. 
%The Appendix contains the formulas necessary for the transition to the coordinate representation.

We use the system of units $ e=\hbar=c=k_B=1 $, and the metric signature $(+,-,-,-)$. The metric in the general case will be denoted by $ g_{\mu\nu} $, and the Minkowski metric - by $ \eta_{\mu\nu} $. Quantum operators are denoted by the hat, e.g. $\, \hat{O} \,$, with the exception of field operators ($ \varphi, \psi, A_{\mu} $, ...).
%================
\section{Theoretical framework and methods}
\label{sec_kubo}
%================

\subsection{Rindler coordinates, Unruh effect and membrane paradigm}

The Rindler coordinates have the well-known form \cite{Crispino:2007eb}
\begin{eqnarray}
ds^2= \rho^2 d\tau^2-d\text{x}^2-d\text{y}^2-
d\rho^2\,,
\label{metricR}
\end{eqnarray}
where the connection with the Minkowski coordinates $ (t,\text{x},\text{y},\text{z}) $ is described by the hyperbolic functions $ z=\rho  \cosh{\tau} $ and $ t=\rho  \sinh{\tau} $. The spacetime with metric (\ref{metricR}) is characterized by the event horizon at the origin of the $ \rho $ coordinate
\begin{eqnarray}
\text{Horizon}: \quad g_{00}(\rho = 0)= 0\,.
\label{hor}
\end{eqnarray}
Let's shift the horizon from the point $ \rho = 0 $ to $ \rho = l_c $ and define the corresponding quantum field theory on the ray $ \rho \in [l_c,\infty) $. The surface $ \rho = 0 $ corresponds to a ``true horizon'', and $ \rho = l_c $ to a ``stretched'' one. Thus, the horizon can be considered as a membrane of finite thickness
\begin{eqnarray}
\text{Membrane}:\quad 0\leqslant\rho\leqslant l_c\,.
\label{membrane}
\end{eqnarray}
The existence of such a membrane is a basic assertion of the famous membrane paradigm \cite{Damour:1978cg, Parikh:1997ma, Thorne:1986iy}. Let us briefly recall the essence of the membrane paradigm, considering a real black hole as an example. The action for space with a black hole looks like \cite{Parikh:1997ma}\footnote{Unlike the rest of the text, in the formula (\ref{parikh}) the definitions from \cite{Parikh:1997ma} with the signature $ (-,+,+,+) $ and Newton's constant $ G_N=1 $ are used.}
\begin{eqnarray} \label{parikh}
S = \frac{1}{16 \pi} \int d^4 x \sqrt{-g} R +
\frac{1}{8 \pi} \int d^3 x \sqrt{\pm h} K + S_{\text{\,matter}},
\label{wil}
\end{eqnarray}
where the first term with scalar curvature $ R $ corresponds to the Einstein-Hilbert action, the third - to the matter living above the stretched horizon, and the second - to the surface term containing the integral over the outer boundary of spacetime (not the stretched horizon) from the trace of the  extrinsic curvature $ K $. By examining the classical equations of motion, considering the boundary terms (interior and outer) when varying, taking into account the cutoff (\ref{membrane}), it can be shown that the black hole membrane has hydrodynamic properties and obeys the Navier-Stokes equation. At the same time, this membrane has a nonzero viscosity, which satisfies the bound (\ref{main}). However, bulk viscosity turns out to be negative in the case of a real black hole $ \zeta<0 $. One can also study the membrane associated with the Rindler horizon \cite{Eling:2009sj, Chirco:2010xx}.

In the current paper, we will be interested not in the membrane itself, but in the quantum fluid living above a stretched horizon, which, actually, corresponds to the third term $ S_{\text{\,matter}} $ in (\ref{wil}). The key in this case is the well-known Unruh effect \cite{Unruh:1976db, Crispino:2007eb}. Let us consider the world line of an observer at rest in Rindler coordinates with constant $ \rho,x,y= \text{const}$. The proper acceleration and time for this observer are described by the Rindler coordinates
\begin{eqnarray}
|a|= \rho^{-1},\, \tau_{\text{\,proper}}=\rho  \tau\,, 
\label{proper}
\end{eqnarray}
where $|a|=\sqrt{-a^{\mu}a_{\mu}}$ is the modulus of the proper acceleration $a_{\mu}=u^{\nu}\nabla_{\nu}u_{\mu}$, and $ u_{\mu} $ is the four-velocity of the observer.
This accelerated observer perceives the Minkowski vacuum as a medium with a temperature
\begin{eqnarray} 
T_U = \frac{|a|}{2 \pi} \,,
\label{TU}
\end{eqnarray}
which is the essence of the famous Unruh effect \cite{Unruh:1976db, Crispino:2007eb}, which is an analogue of the Hawking effect in the absence of real space-time curvature (that is, it corresponds to the limit of vanishing Newtonian constant $ G_{\text{N}} \to 0 $) . Thus, we can talk about a medium, or thermal bath of particles above the Rindler horizon (similar to Hawking radiation) even for a Minkowski vacuum state. 

Our goal is to study the hydrodynamic properties of this thermal bath, which, as we will see, significantly depend on the loop quantum field effects.

\subsection{Kubo formulas for viscocity: application to the Rindler space}

Let us consider a relativistic viscous fluid. One of the main quantities describing the hydrodynamic properties of such a medium is the energy-momentum tensor (EMT), which looks like the sum of the EMT of an ideal fluid and dissipative correction \cite{LandauFluid}
\begin{eqnarray}
T_{\mu\nu} &=& T_{\mu\nu}^{\,\text{ideal}}+ T_{\mu\nu}^{\text{\,diss}}\,.
\label{TidDiss}
\end{eqnarray}
The EMT of an ideal fluid with velocity $ u_{mu} $ is expressed in terms of energy density $ \varepsilon $ and pressure $ p $
\begin{eqnarray}
T_{\mu\nu}^{\,\text{ideal}} = (\varepsilon+p) u_{\mu}u_{\nu} - p g_{\mu\nu}\,.
\label{TidDiss}
\end{eqnarray}
The form of the dissipative part is fixed by the second law of thermodynamics
\begin{eqnarray}
T_{\mu\nu}^{\,\text diss} &=& -\eta (\nabla_{\mu} u_{\nu}+ \nabla_{\nu} u_{\mu} -u_{\mu} u^{\alpha}\nabla_{\alpha}u_{\nu} -u_{\nu} u^{\alpha}\nabla_{\alpha}u_{\mu} ) -\left( \zeta -\frac{2}{3}\eta\right)\nabla^{\alpha}u_{\alpha}(g_{\mu\nu}-u_{\mu}u_{\nu}) + \mathcal{O}(\nabla^2 u) \,,
\label{Tdiss}
\end{eqnarray}
where $ \eta \geqslant 0 $ is the shear viscosity, and $ \zeta \geqslant 0 $ is the bulk viscosity, $ u_{\mu} $ is the four-velocity of the fluid, and $ \nabla_{\mu} $ is in general case a covariant derivative. $ \mathcal{O}(\nabla^2 u) $ denotes terms of higher orders in gradients (for example, containing $ \nabla_{\mu} u_{\nu} \nabla_{\alpha} u_{\beta} $). The non-negativity of $\eta$ and $\zeta$ follows from the second law of thermodynamics \cite{LandauFluid} and ensures thermodynamic stability \cite{Kovtun:2019hdm}.

In this paper, we are primarily interested in shear viscosity $ \eta $. Transport coefficients in the general case can be calculated using the linear response theory \cite{zubarev1974nonequilibrium, Laine:2016hma, Son:2008zz}. The corresponding formulas are called Kubo formulas. Shear viscosity can be calculated using the Kubo formula containing the correlator of two EMT operators averaged over a certain quantum state
\begin{eqnarray}
\eta = \lim_{\omega \to 0} \frac{1}{\omega} \int d^4 x e^{i\omega t} \theta(t)\langle [\hat{T}_{\text{xy}}(x),\hat{T}_{\text{xy}} (0)] \rangle \,,
\label{kuboInit}
\end{eqnarray}
where $ \theta(t) $ is the Heaviside step function, and $ \hat{T}_{\text{xy}} $ is the off-diagonal component of the energy-momentum tensor. The appearance of two EMT operators can be understood, for example, if we consider that the effects of the velocity derivatives $ \partial_{\mu} u_{\mu} $ can be obtained from the interaction vertex with the gravitational field $ \delta g_{\mu\nu } \hat{T}^{\mu\nu} $ (see, for example, \cite{Son:2008zz}). We also note that according to (\ref{kuboInit}), the shear viscosity $ \eta $ is described by the Fourier transform of a given correlator, which contains, actually, the double limit $ \omega,\vec{q} \to 0 $, and initially tends to zero the spatial part of the momentum $ \vec{q} \to 0 $. This arrangement of limits reflects the dissipative nature of the corresponding transport coefficient \cite{Landsteiner:2012kd}.

Passing to the Rindler space (\ref{metricR}), we obtain from (\ref{kuboInit}), according to \cite{Chirco:2010xx}
\begin{eqnarray}\label{kuboRindler}
\eta=\pi \lim_{\omega\rightarrow0}{\int_{l_c}^{\infty}\rho'\,d\rho'\int_{l_c}^{\infty}\rho\,d\rho\int_{-\infty}^{\infty}\,d\text{x}\,d\text{y}\,d\tau e^{i\omega\tau}\langle 0|\hat{T}_{\text{xy}}(\tau, \text{x}, \text{y}, \rho)\hat{T}_{\text{xy}}(0, 0, 0, \rho')|0 \rangle_{\text{M}}} \,,
\end{eqnarray}
where the integrals over $ d\rho $ and $ d\rho' $ take into account that the fields do not penetrate inside the membrane. The last integral $ \int d\rho' $ is due to the fact that viscosity is calculated per unit area of the horizon. We are interested in the viscosity for the Minkowski vacuum $ |0 \rangle_{\text{M}} $, that is, when $ T=T_U $. Note that (\ref{kuboRindler}), unlike (\ref{kuboInit}) (which contains a retarded Green's function), contains a Wightman function (see, for example, \cite{Bogoliubov1983-di}), the transition to which is possible in the limit $ \omega \to 0 $ (for more details, see \cite{Chirco:2010xx}).

We will call the quantity obtained before the last integration $ \int d\rho' $  in (\ref{kuboRindler}) \textit{local} viscosity $ \eta_{\text{\,loc}}(\rho') $
\begin{eqnarray} \label{etaLocDef}
\eta = \int_{l_c}^{\infty}\,d\rho' \, \eta_{\text{\,loc}}(\rho') 
\,.
\end{eqnarray}
and the integrated quantity  as \textit{global} viscosity. We will return to discuss this in more detail in Section \ref{subsec_local}. 

\begin{figure*}[!h]
\begin{minipage}{0.4\textwidth}
  \centerline{\includegraphics[width=1\textwidth]{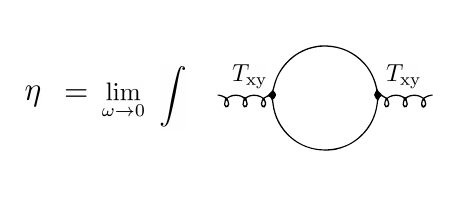}}
\end{minipage}
\caption{Feynman diagram corresponding to the calculation of shear viscosity using the Kubo formula (\ref{kuboRindler}).}
\label{fig:1}
\end{figure*}

We will consider free fields, but as we will see, even for free fields viscosity arises (see discussion in Section \ref{subsec_unruh}). So, a typical correlator that we will calculate using (\ref{kuboRindler}), has the form, shown in Figure \ref{fig:1}.

%Следуя \cite{Chirco:2010xx}, вместо коммутатора в () можно рассмотреть коррелятор вида ()
%
%так как ()
%
%и запаздывающая функция Грина не отличается от () в рассматриваемом нами Соответствующая диаграмма для коррелятора (), задающего вязкость, приведена на Рисунке ().

\subsection{Entropy in the Rindler space}

To calculate the ratio (\ref{nsmain}), we also need to know the entanglement entropy density in Rindler space. We will follow the approach used in particular in \cite{Becattini:2023ouz}, according to which entropy density can be calculated as the derivative of pressure at constant acceleration
\begin{eqnarray}\label{sDef}
s_{\text{\,loc}}=\pdv{p}{T}\Big|_{a} \,,
\end{eqnarray}
where the subscript $loc$ reflects that the quantity corresponds to a point at a certain distance from the horizon. It is important that (\ref{sDef}) contains the derivative at constant acceleration $ a_{\mu}=\text{const} $. Therefore, as we will show later, the $ |a|^2 T^2 $ term from pressure contributes to the entropy density for spins higher than $ 0 $. We will discuss this definition of entropy in more detail in Section \ref{subsec_entropy}.

The corresponding energy density and pressure can be found from the mean value of the EMT operator
\begin{eqnarray}\label{pDef}
\varepsilon= \langle \hat{T}_{\mu\nu} \rangle u^{\mu}u^{\nu}\,,\quad
p= - \frac{1 }{3}\langle \hat{T}_{\mu\nu} \rangle \Delta^{\mu\nu}\,,
\end{eqnarray}
where the projector operator is introduced $ \Delta_{\mu\nu} = g_{\mu\nu} - u_{\mu} u_{\nu} $. Since we are considering the Minkowsky vacuum, we can put after calculating the derivative in (\ref{sDef}) $ T=T_U=|a|/2\pi $. If we then take into account that $ |a|=1/\rho $, according to (\ref{proper}), then the entropy density will be a function only of the distance to the horizon
\begin{eqnarray}\label{sDef1}
\text{Minkowsky vacuum:}\quad s_{\text{\,loc}}(T=T_U,|a|)\underrel{|a|\to1/\rho}{=}s_{\text{\,loc}}(\rho) \,.
\end{eqnarray}

We will also be interested in the total enteropy per unit area of the horizon and therefore we will integrate it over $ \rho $ from $ l_c $ to infinity
\begin{eqnarray}\label{sDef2}
s=\int_{l_c}^{\infty} \, d\rho\,   s_{\text{\,loc}}(\rho) \,.
\end{eqnarray}

Let us proceed directly to calculating viscosity and entropy for various field theories in Rindler space.

%================
\section{Shear viscosity for scalar field}
\label{sec_scalar}
%================

\subsection{Shear viscosity calculation}

The case of a massless scalar field was considered in \cite{Chirco:2010xx}. We will briefly describe the corresponding derivation for viscosity, moving the more detailed description of the calculation to the next section, where we consider Dirac fields.

Let us consider the general case of the improved energy-momentum tensor of the real massless scalar field $ \varphi $ in four dimensions in flat space-time \cite{Birrell:1982ix}
\begin{eqnarray} \label{Tscalar}
T_{\mu\nu} = (1-2\xi)\partial_{\mu} \varphi \partial_{\nu} \varphi
+(2\xi - \frac{1}{2})\eta_{\mu\nu} \partial_{\alpha} \varphi  \partial^{\alpha} \varphi 
-2 \xi (\partial_{\mu}\partial_{\nu} \varphi) \varphi
+ \frac{\xi}{2} \eta_{\mu\nu} \varphi  \partial^{\alpha}\partial_{\alpha} \varphi\,,
\end{eqnarray}
where conformal symmetry is achieved at $ \xi=1/6 $. Note that since we are interested in the quantum average over the Minkowski vacuum, the calculation can be performed in Minkowski coordinates with the metric $ \eta_{\mu\nu} $ and go to the Rindler coordinates only before finding the Fourier transform of the matrix element.

Correlator $ \langle 0 | \hat{T}_{\mu\nu} (x) \hat{T}_{\alpha\beta} (y)|0\rangle_{\text{M}} $ can be found using a scalar field propagator in the coordinate representation, which is taken as the Wightman function. Since the theory is massless, the calculation is reduced to a simple multiplication of propagators and vertices. As a result, we obtain
\begin{eqnarray} \label{TTscalar}
\langle 0 | \hat{T}_{\mu\nu} (x) \hat{T}_{\alpha\beta} (y)|0\rangle_{\text{M}} =
\frac{4}{3\pi^4} \mathcal{I}_{\mu\nu\alpha\beta}(x-y)+\frac{240(\xi-1/6)^2}{\pi^4} \widetilde{\mathcal{I}}_{\mu\nu\alpha\beta}(x-y)\,,
\end{eqnarray}
where the first term corresponds to the universal two-point function with two EMT operators in conformal field theory \cite{Erdmenger:1996yc}
\begin{eqnarray} \label{I}
\mathcal{I}_{\mu\nu\alpha\beta} (b) = \frac{b_{\mu} b_{\nu} b_{\alpha} b_{\beta}}{\bar{b}^{12}}
-\frac{\eta_{\mu\alpha} b_{\nu} b_{\beta}}{4\bar{b}^{10}}
-\frac{\eta_{\nu\alpha} b_{\mu} b_{\beta}}{4\bar{b}^{10}}
-\frac{\eta_{\mu\beta} b_{\nu} b_{\alpha}}{4\bar{b}^{10}}
-\frac{\eta_{\nu\beta} b_{\mu} b_{\alpha}}{4\bar{b}^{10}}
+\frac{\eta_{\mu\alpha} \eta_{\nu \beta}}{8\bar{b}^8}
+\frac{\eta_{\mu\beta} \eta_{\nu \alpha}}{8\bar{b}^8}
-\frac{\eta_{\mu\nu} \eta_{\alpha \beta}}{16\bar{b}^8}\,,
\end{eqnarray}
where 
\begin{eqnarray} \label{poles}
b_{\mu} = x_{\mu}-y_{\mu}\,, \quad \bar{b}^2 = b^2 - i \varepsilon b_0\,,
\end{eqnarray}
and $ \varepsilon>0, \varepsilon \to 0 $. The second term in (\ref{TTscalar}) describes the deviation from the conformally symmetric form
\begin{eqnarray} \nonumber
\widetilde{\mathcal{I}}_{\mu\nu\alpha\beta} (b) &=& \frac{b_{\mu} b_{\nu} b_{\alpha} b_{\beta}}{\bar{b}^{12}}
-\frac{\eta_{\mu\alpha} b_{\nu} b_{\beta}}{10 \bar{b}^{10}}
-\frac{\eta_{\nu\alpha} b_{\mu} b_{\beta}}{10 \bar{b}^{10}}
-\frac{\eta_{\mu\beta} b_{\nu} b_{\alpha}}{10 \bar{b}^{10}}
-\frac{\eta_{\nu\beta} b_{\mu} b_{\alpha}}{10 \bar{b}^{10}}
+\frac{\eta_{\mu\alpha} \eta_{\nu \beta}}{80 \bar{b}^8}
+\frac{\eta_{\mu\beta} \eta_{\nu \alpha}}{80 \bar{b}^8}
+\frac{13\eta_{\mu\nu} \eta_{\alpha \beta}}{80 \bar{b}^8} \\
&& -\frac{3\eta_{\mu\nu} b_{\alpha} b_{\beta}}{10 \bar{b}^{10}}
-\frac{3\eta_{\alpha\beta} b_{\mu} b_{\nu}}{10 \bar{b}^{10}}\,.
\label{Itilde}
\end{eqnarray}
Note that the form of (\ref{I}) and (\ref{Itilde}) up to coefficients in front of each term is in an obvious way can be fixed from the dimensional and symmetry considerations. In accordance with the choice of the positive-frequency Wightman function as the Green's function, the poles in (\ref{I}) and (\ref{Itilde}) are shifted upward from the real time axis.

It is interesting to note that the dependence on the conformal symmetry parameter $ \xi $ goes away after the integration in the horizon plane contained in the formula (\ref{kuboRindler})
\begin{eqnarray} \label{IntHorScalar}
\int d\text{x}\, d\text{y}\, \langle 0 | \hat{T}_{\mu\nu} (t,\text{x},\text{y},\text{z}) \hat{T}_{\alpha\beta} (0,0,0,\text{z}')|0\rangle_{\text{M}} = -\frac{1}{30 \pi^3 (t^2-(\text{z}-\text{z}')^2-i\varepsilon t)^3}\,,
\end{eqnarray}
which can be done explicitly by passing to polar coordinates. As a result, the following expression for ``local viscosity'' (\ref{etaLocDef}) was obtained 
\begin{eqnarray} \label{etaScalarLoc}
		\begin{split}
			\eta^{\text{\,scalar}}_{\text{\,loc}} (\rho)=\frac{\rho\left[\rho^4 + 4\rho^2 l_c^2 - 5l_c^4 - 
				4 l_c^2 (2 \rho^2+l_c^2) \ln\frac{\rho}{l_c}\right]}{240 (\rho^2-l_c^2 )^4 \pi^2}\,.
		\end{split}
\end{eqnarray}
The total viscosity (\ref{etaLocDef}) per unit area of the horizon, will be equal to
\begin{eqnarray} \label{etaScalar}
\eta^{\text{\,scalar}} = \frac{1}{1440 \pi^2 l_c^2}
\,.
\end{eqnarray}
The divergence when approaching the horizon at $ l_c\to 0 $ is typical for Rindler space, as we will see in the next subsection.

\subsection{Entropy calculation}

Let us now turn to entropy. The mean value of the EMT of a gas of massless real scalar field with temperature $ T $ and acceleration $ a_{\mu} $ is well known \cite{Dowker:1994fi, Zakharov:2020ked}\footnote{Temperature and acceleration are two independent parameters and point $ T=|a|/2\pi $ corresponds to a concrete choice of the quantum state of the system in the form of a Minkowski vacuum. Note also that, unlike (\ref{TU}), acceleration $ a_{\mu} $ and velocity $ u_{\mu} $ in (\ref{EMTscalar}) are related to the medium, so the medium acts as a detector. In particular, this is evident in the equality $ \langle \hat{T}_{\mu\nu}\rangle (T=T_U)=0 $, for details see \cite{Becattini:2017ljh}.}
\begin{eqnarray} \label{EMTscalar}
\langle \hat{T}^{\text{\,scalar}}_{\mu\nu}\rangle = \Big(\frac{\pi^2 T^4}{30}-\frac{|a|^4}{480 \pi^2}\Big)\Big(u_{\mu}u_{\nu}-\frac{\Delta_{\mu\nu}}{3}\Big)
\,.
\end{eqnarray}
Note that we are limited to the conformally symmetric case $ \xi=1/6 $, since for arbitrary $ \xi $ the energy-momentum tensor becomes anisotropic \cite{Zakharov:2020ked} and it is not entirely clear how to find the entropy. The pressure in the fluid rest frame, according to (\ref{pDef}) and (\ref{EMTscalar}), has the form
\begin{eqnarray} \label{pScalar}
p^{\text{\,scalar}} (T,a) =\frac{1}{3} \Big(\frac{\pi^2 T^4}{30}-\frac{|a|^4}{480 \pi^2}\Big)
\,.
\end{eqnarray}
According to (\ref{sDef}), the entropy density can be found as a derivative at constant acceleration. We obtain (in accordance with \cite{Becattini:2023ouz})
\begin{eqnarray}\label{sScalarAT}
s^{\text{\,scalar}}_{\text{\,loc}}(T)=\frac{2 \pi^2 T^3}{45} \,.
\end{eqnarray}
Note that the vacuum term $ |a|^4 $ from (\ref{pScalar}) does not contribute to the entropy. For the Minkowski vacuum, according to (\ref{sDef1}), we obtain
\begin{eqnarray}\label{sScalarLoc}
s^{\text{\,scalar}}_{\text{\,loc}} (\rho)=\frac{1}{180 \pi \rho^3} \,.
\end{eqnarray}
The total entropy per unit area of the horizon according to (\ref{sDef2}) will have the form
\begin{eqnarray} \label{sScalar}
s^{\text{\,scalar}}=\frac{1}{360\pi l_c^2}\,.
\end{eqnarray}

\subsection{Shear viscosity to entropy ratio}

Note that both the entropy $ s $ and the viscosity $ \eta $ in (\ref{sScalar}) and (\ref{etaScalar}) diverge in the limit $ l_c\to \infty $, which is typical for the Rindler space. However, their ratio is finite and independent of $l_c$
\begin{eqnarray} \label{etaSglobScalar}
\frac{\eta}{s}\Big|_{\text{scalar}}=\frac{1}{4\pi}\,.
\end{eqnarray}
Thus, in the case of scalar fields, the Minkowski vacuum corresponds to the minimum ratio of viscosity to entropy (\ref{main}).

The ratio of local viscosity (\ref{etaScalarLoc}) to local entropy (\ref{sScalarLoc}) is a function of the distance from the horizon and depends on the thickness of the membrane $ l_c $
\begin{eqnarray} \label{etaSloc}
\frac{\eta_{\text{\,loc}}}{s_{\text{\,loc}}}(\rho)= f(\rho/l_c) =
\frac{3\rho^4
\left[\rho^4 +4\rho^2{l_c}^2-5{l_c}^4-4{l_c}^2(2\rho^2+l_c^2)\ln\left(\frac{\rho}{l_c}\right)
\right]}{4\pi(\rho^2-l_c^2)^4} \,,
\end{eqnarray}
where
\begin{eqnarray} \label{f}
f(x)=\frac{x^4 (x^4+4x^2-5-4(2x^2+1)\ln x)}{4 \pi (x-1)^4} \,.
\end{eqnarray}
In the next two sections we will show that many of the discovered properties are universal for particles with higher spins.

%================
\section{Shear viscosity for Dirac field}
\label{sec_dirac}
%================

\subsection{Shear viscosity calculation}

Let us now consider Dirac fields, also limiting ourselves to the massless case and using the same algorithm as in the case of a scalar field. The Belinfante EMT for massless Dirac fields has the form
\begin{eqnarray}\label{TDirac}
T_{\mu\nu}=\frac{i}{4}(\bar{\psi} \gamma_{\mu}\partial_{\nu} \psi-\partial_{\nu}\bar{\psi} \gamma_{\mu} \psi+\bar{\psi} \gamma_{\nu}\partial_{\mu} \psi- \partial_{\mu}\bar{\psi} \gamma_{\nu}\partial_{\nu} \psi)\,.
\end{eqnarray}
As before, we will take advantage of the fact that we are interested in the corresponding matrix element over the Minkowski vacuum and calculate it in Minkowski coordinates, moving to Rindler coordinates before taking the Fourier transform. The propagator in the coordinate representation has the form \cite{Bogoliubov1983-di} \footnote{One of the simple ways to obtain the Wightman function in the coordinate representation is to consider the well-known causal Feynman propagators, integrate over the momentum (for example, using Eqs. from \cite{Grozin:2003ak}) and use the connection between the various Green functions.}
\begin{eqnarray}\label{propDiracX}
S_{ab}(x)=\langle 0| 
\psi_{a}(x)\bar{\psi}_{b}(0)
|0\rangle_{M} =
 \frac{i}{2 \pi^2} \frac{(\gamma x)_{ab}}{(x^2-i\varepsilon x_0)^2} \,,
\end{eqnarray}
where $ a, b $ are bispinor indices and the numerator contains convolution with Dirac matrices $ (\gamma p) = \gamma_{\mu}p^{\mu} $. The Green function (\ref{propDiracX}) is the so-called Wightman function. The choice of such a Green function is due to the fact that (\ref{kuboRindler}) also contains the Wightman function of two EMT operators. Note that the poles in (\ref{propDiracX}) are shifted upward relative to the real time axis $ \Re t $.
%As usual in the case of conformal field theory and massless integrals, we go to the coordinate representation, using the formula (\ref{A3}) from the Appendix
%\begin{eqnarray}\label{propDiracP}
%S_{ab}(x)=\langle 0| 
%\psi_{a}(x)\bar{\psi}_{b}(0)
%|0\rangle_{M} =
%i \int \frac{d^4 p}{(2 \pi)^4} e^{-ipx} \frac{(\gamma p)_{ab}}{p^2}
%   \,,
%\end{eqnarray}

It is convenient to represent EMT in (\ref{TDirac}) (quantum operator) in split form
\begin{eqnarray} \nonumber 
\hat{T}_{\mu\nu}(x) &=& \lim_{x_1,x_2\to x} \mathcal{D}^{ab}_{\mu\nu }(\partial_{x_1},\partial_{x_2})\bar{\psi}_{a}(x_1)\psi_{ b}(x_2)\,, \\
\mathcal{D}^{ab}_{\mu\nu }(\partial_{x_1},\partial_{x_2}) &=& \frac{i}{4}
(\gamma^{ab}_{\mu}\partial_{\nu}^{x_2}-\gamma^{ab}_{\mu}\partial_{\nu}^{x_1}+
\gamma^{ab}_{\nu}\partial_{\mu}^{x_2}-\gamma^{ab}_{\nu}\partial_{\mu}^{x_1})\,, \label{TDdirac}
\end{eqnarray}
which is not regularization in this case, but is only necessary to take the derivatives out of the bracket. Substituting (\ref{TDdirac}) into (\ref{kuboInit}), we obtain
\begin{eqnarray} \label{TTdiracSplit}
\langle 0 | \hat{T}_{\mu\nu} (x) \hat{T}_{\alpha\beta} (y)|0\rangle_{\text{M}} = \lim_{
	\def\arraystretch{0.5}\begin{array}{ll}
	{\scriptscriptstyle x_1,x_2\to x}\vspace{0.1mm}\\
	{\scriptscriptstyle y_1,y_2\to y}
	\end{array}}
\mathcal{D}^{ab}_{\mu\nu }(\partial_{x_1},\partial_{x_2})
\mathcal{D}^{cd}_{\alpha\beta }(\partial_{y_1},\partial_{y_2})
\langle 0 | \bar{\psi}_{a}(x_1)\psi_{b}(x_2) \bar{\psi}_{c}(y_1)\psi_{d}(y_2)|0\rangle_{\text{M}}
\,.
\end{eqnarray}
Using Wick's theorem \cite{Bogoliubov1983-di} and limiting ourselves to considering only connected contributions, we obtain \footnote{We use that $ \langle 0| 
\bar{\psi}_{a}(x)\psi_{b}(0)
|0\rangle_{M} =  S_{ba} (x) $.}
\begin{eqnarray} \nonumber
\langle 0 | \bar{\psi}_{a}(x_1)\psi_{b}(x_2) \bar{\psi}_{c}(y_1)\psi_{d}(y_2)|0\rangle_{M,\text{connected}} &=&
\langle 0 | \bar{\psi}_{a}(x_1)\psi_{d}(y_2)|0\rangle_{\text{M}} \langle 0 | \psi_{b}(x_2) \bar{\psi}_{c}(y_1) |0\rangle_{\text{M}} \\
&=&  S_{da}(x_1-y_2)S_{bc}(x_2-y_1)
\,. \label{wickDirac}
\end{eqnarray}
Substituting (\ref{wickDirac}) into (\ref{TTdiracSplit}), we will have
\begin{eqnarray} \nonumber
\langle 0 | \hat{T}_{\mu\nu} (x) \hat{T}_{\alpha\beta} (y)|0\rangle_{\text{M}}  &=&
\frac{1}{16} \text{tr}\Big\{ 
\gamma_{\mu} \partial_{\nu} S(b) \gamma_{\alpha} \partial_{\beta} S(b)
-\gamma_{\mu} \partial_{\beta}\partial_{\nu} S(b) \gamma_{\alpha}  S(b)
+\gamma_{\mu} \partial_{\nu} S(b) \gamma_{\beta} \partial_{\alpha} S(b) \nonumber
\\ &&-\gamma_{\mu} \partial_{\alpha}\partial_{\nu} S(b) \gamma_{\beta}  S(b)
- \gamma_{\mu} S(b) \gamma_{\alpha} \partial_{\nu} \partial_{\beta} S(b)
+\gamma_{\mu}\partial_{\beta}  S(b) \gamma_{\alpha} \partial_{\nu} S(b) \nonumber
\\ &&-\gamma_{\mu}  S(b) \gamma_{\beta} \partial_{\alpha}\partial_{\nu} S(b)
+\gamma_{\mu} \partial_{\alpha} S(b) \gamma_{\beta} \partial_{\nu} S(b)
+\gamma_{\nu} \partial_{\mu} S(b) \gamma_{\alpha} \partial_{\beta} S(b) \nonumber
\\ &&-\gamma_{\nu} \partial_{\beta}\partial_{\mu} S(b) \gamma_{\alpha}  S(b)
+\gamma_{\nu} \partial_{\mu} S(b) \gamma_{\beta} \partial_{\alpha} S(b)
-\gamma_{\nu} \partial_{\alpha}\partial_{\mu} S(b) \gamma_{\beta}  S(b) \nonumber
\\ &&-\gamma_{\nu} S(b) \gamma_{\alpha} \partial_{\beta} \partial_{\mu} S(b)
+\gamma_{\nu} \partial_{\beta} S(b) \gamma_{\alpha} \partial_{\mu} S(b)
-\gamma_{\nu} S(b) \gamma_{\beta} \partial_{\alpha}  \partial_{\mu} S(b) \nonumber
\\ &&+\gamma_{\nu} \partial_{\alpha} S(b) \gamma_{\beta} \partial_{\mu} S(b)
\Big\} \,, \label{TTdiracInterm}
\end{eqnarray}
where we have again introduced the vector $ b_{\mu} = x_{\mu}-y_{\mu} $, and the derivatives are of the form $ \partial_{\mu} \equiv \partial/\partial b^{\mu} $. Substituting (\ref{propDiracX}) into (\ref{TTdiracInterm}), differentiating and taking a trace, we finally obtain
\begin{eqnarray}\label{TTdirac}
\langle 0 | \hat{T}_{\mu\nu} (x) \hat{T}_{\alpha\beta} (y)|0\rangle_{\text{M}}  = \frac{8}{\pi^4} \mathcal{I}_{\mu\nu\alpha\beta} (x-y)
\,,
\end{eqnarray}
where $ \mathcal{I} $ is defined in (\ref{I}).

\begin{figure*}[!h]
\begin{minipage}{0.7\textwidth}
  \centerline{\includegraphics[width=1\textwidth]{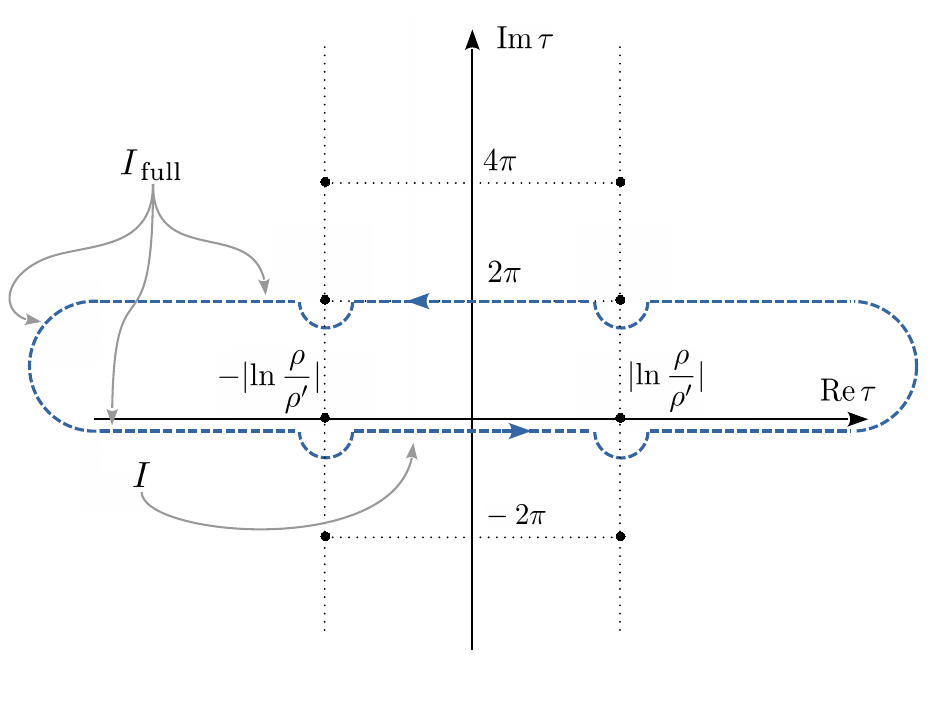}}
\end{minipage}
\caption{Calculation of the integral (\ref{intIformula}) using Cauchy's theorem. The integral $ I $ is calculated along the real axis, and the integral $ I_{\text{full}} $ is along the closed contour (dashed blue line). The dots show the poles in the integrand in (\ref{intIformula}) taking into account the rule for bypassing them.}
\label{fig:2}
\end{figure*}

Thus, we have found the necessary correlator and all that remains is to find its Fourier transform. We will now substitute (\ref{TTdirac}) into (\ref{kuboRindler}). It is convenient to integrate in the horizon plane $ (\text{x},\text{y}) $ by passing to polar coordinates
\begin{eqnarray}\label{polar}
\text{x}= r \cos{\phi}\,,\quad \text{y}=r \sin{\phi}
\,.
\end{eqnarray}
After this we obtain
\begin{eqnarray}\label{inthorDiracFormula}
\int_{-\infty}^{\infty}\int_{-\infty}^{\infty} d\text{x}\, d\text{y}\, \langle 0 | \hat{T}_{\text{xy}} (t,\text{x},\text{y},\text{z}) \hat{T}_{\text{xy}} (0,0,0,\text{z}')|0\rangle_{\text{M}}  = \int_{0}^{\infty} r dr \int_{0}^{2\pi} d\phi \, \frac{1}{\pi^4} \frac{{\alpha^2 - r^4 \cos(4 \phi)}}{(\alpha + r^2)^6}
\,,
\end{eqnarray}
where the designation is introduced $ \alpha = -t^2 + (z-z')^2 + i\varepsilon t $. The integral over the angle $ \phi $ can be taken in a trivial way. The integral over $ r $ can be found directly, since due to the infinitesimal shift by an imaginary value in the denominator, the integral over the real axis $ r $ does not contain the poles
\begin{eqnarray} \label{inthorDiracResult}
\int_{0}^{\infty} r dr \int_{0}^{2\pi} d\phi \, \langle 0 | \hat{T}_{\text{xy}} \hat{T}_{\text{xy}}|0\rangle_{\text{M}} = \frac{1}{5 \pi^3 \alpha^3} \,.
\end{eqnarray}
The next step is to integrate over the Rindler time $ \tau $ in (\ref{kuboRindler}). To do this, let us pass in the corresponding integral to the Rindler coordinates
\begin{eqnarray} \label{intIformula}
I=\pi \int_{-\infty}^{\infty} d\tau e^{i \tau \omega} \frac{1}{5 \pi^3 \alpha^3} = 
\int_{-\infty}^{\infty}\frac{ e^{i \tau \omega}}{5 \pi^2 (\rho^2 + \rho'^2 - 2 \rho \rho' \cosh(\tau) + i \varepsilon \tau)^3} \,d\tau \,.
\end{eqnarray}
The integrand contains poles at points located along two vertical lines
\begin{eqnarray} \label{polesTau}
\tau = \pm \ln{\frac{\rho}{\rho'}}(1 + i \varepsilon) +2 \pi i n \quad n=0,\pm 1,\pm 2 ...\,,
\end{eqnarray}
where the rule for bypassing the poles follows from the form determined by the infinitesimal shift $ i \varepsilon \tau $. The corresponding poles are shown in Figure \ref{fig:2}. Let's close the contour by shifting $ \tau \to \tau +2 \pi i $. Since $ \cosh (\tau + 2 \pi i)= \cosh (\tau)$ and $ \exp[i \omega (\tau + 2 \pi i)] =\exp(i \omega \tau) \exp(-2 \pi \omega) $, we can relate this contour integral $ I_{\text{full}} $ with the integral $ I $ of interest to us
\begin{eqnarray} \label{IIfull}
I=(1-e^{-2\pi \omega})^{-1} I_{\text{full}} \,.
\end{eqnarray}

The integral (\ref{intIformula}) can be found using Cauchy's theorem. Only two poles $ \tau = \pm \ln{\frac{\rho}{\rho'}} $ fall inside the contour, as shown in Figure \ref{fig:2}
\begin{eqnarray} \label{intIfull}
I_{\text{full}} = 2\pi i \sum_{\tau_0 = \pm \ln{\frac{\rho}{\rho'}} } \underset{\tau \to \tau_0}{\text{Res}}\, \frac{e^{i \tau \omega}}{5 \pi^2 [\rho^2 + \rho'^2 - 2 \rho \rho' \cosh(\tau)]^3}\,.
\end{eqnarray}
As a result, we obtain
\begin{eqnarray} \label{intIresult}
I = \frac{-6 \omega \left(\rho^4-\rho'^4\right) 
\cos{\left(  \omega \ln\frac{\rho}{\rho'}\right)} -
2\left\{(\rho^4 + \rho'^4) (2-\omega^2) + 2\rho^2\rho'^2(4+\omega^2)\right\}
\sin{\left(\omega \ln\frac{\rho}{\rho'}\right)}}{5 \pi (\rho^2-{\rho'}^2)^5(1- e^{-2 \pi \omega})}\,,
\end{eqnarray}
According to (\ref{kuboRindler}) it is necessary to go to the limit $ \omega \to 0 $
\begin{eqnarray} \label{intWlim}
\lim_{\omega \to 0} I = \frac{3\rho'^4-3 \rho^4 +2
[\rho^4+4\rho^2\rho'^2+\rho'^4]\ln\frac{\rho}{\rho'}}{5\pi^2 (\rho^2-\rho'^2)^5}\,.
\end{eqnarray}
Note that in (\ref{intWlim}) there was mutual cancellation of the divergences $ 1/\omega $ from the two poles $ \tau_0 = \pm \ln{\frac{\rho}{\rho'}} $. Now, integrating explicitly over $ \rho $ (the integral is convergent), we obtain the following expression for the local viscosity (\ref{etaLocDef}) (we replace $ \rho'\to\rho $)
\begin{eqnarray} \label{etaDiracLoc}
			\eta^{\text{\,Dirac}}_{\text{\,loc}} (\rho)=\frac{\rho\left[\rho^4
+ 4 \rho^2 l_c^2-5l_c^4  - 
				4 l_c^2 (2 \rho^2+l_c^2) \ln \frac{\rho}{l_c} \right]}{40 ( \rho^2-l_c^2)^4 \pi^2}\,.
\end{eqnarray}
To find the global viscosity, it remains to take the integral over $ \rho $ according to (\ref{etaLocDef}), which converges within the chosen limits and can be found directly
\begin{eqnarray} \label{etaDirac}
\eta^{\text{\,Dirac}} =  \frac{1}{240 \pi^2 l_c^2}
\,.
\end{eqnarray}

\subsection{Entropy calculation}

Let's now move on to entropy. The mean value of the EMT for massless Dirac fields at zero chemical potential in Rindler space has the form
\begin{eqnarray} \label{EMTdirac}
\langle \hat{T}^{\text{\,Dirac}}_{\mu\nu}\rangle = \Big(\frac{7\pi^2 T^4}{60}+\frac{T^2 |a|^2}{24}-\frac{17 |a|^4}{960 \pi^2}\Big)\Big(u_{\mu}u_{\nu}-\frac{\Delta_{\mu\nu}}{3}\Big)
\,.
\end{eqnarray}
This expression can be obtained either by quantizing fields in the (Euclidean) Rindler space \cite{Dowker:1994fi, Prokhorov:2019yft, Frolov:1987dz}, or in the usual Minkowski space from statistical interaction with boost generator \cite{Dowker:1994fi, Prokhorov:2019cik, Prokhorov:2019yft, Palermo:2021hlf}\footnote{This correspondence indicates the duality of the quantum-statistical and geometric approaches \cite{Prokhorov:2019yft}.}. However, note that, unlike scalar fields, (\ref{EMTdirac}) is expected to be valid only for $ T>T_U $ due to the phase transition at $ T=T_U $ \cite{Prokhorov:2023dfg, Prokhorov:2019hif}.

The pressure (\ref{pDef}) has the form
\begin{eqnarray} \label{pDirac}
p^{\text{\,Dirac}} (T,a) =\frac{1}{3} \Big(\frac{7\pi^2 T^4}{60}+\frac{T^2 |a|^2}{24}-\frac{17 |a|^4}{960 \pi^2}\Big)
\,.
\end{eqnarray}
Using the definition (\ref{sDef}), we obtain the entropy density
\begin{eqnarray}\label{sLocDiracTa}
s^{\text{\,Dirac}}_{\text{\,loc}}(T,a)=\frac{7\pi^2 T^3}{45} +\frac{T |a|^2}{36} \,.
\end{eqnarray}
For the Minkowsky vacuum state, according to (\ref{sDef1}), we will have
\begin{eqnarray}\label{sDiracLoc}
s^{\text{\,scalar}}_{\text{\,loc}}(T=T_U,a)=
s^{\text{\,scalar}}_{\text{\,loc}}(\rho)= \frac{1}{30 \pi \rho^3} \,.
\end{eqnarray}
The total entropy per unit area of the horizon (\ref{sDef2}) will be
\begin{eqnarray} \label{sDirac}
s^{\text{\,Dirac}} = \frac{1}{60 \pi l_c^2}
\,.
\end{eqnarray}

\subsection{Shear viscosity to entropy ratio}

The ratio of local viscosity (\ref{etaDiracLoc}) and entropy (\ref{sDiracLoc}) for Dirac fields is equal to this ratio for scalar fields and is determined by the same universal function (\ref{etaSloc}), (\ref{f}) 
\begin{eqnarray}
\frac{\eta_{\text{\,loc}}}{s_{\text{\,loc}}}(\rho)\Big|_{\text{Dirac}}= f(\rho/l_c)\,.
\end{eqnarray}
Similarly for global quantities, the ratio of viscosity (\ref{etaDirac}) and entropy (\ref{sDirac}) corresponds to the KSS bound
\begin{eqnarray} \label{etaSglobDirac}
\frac{\eta}{s}\Big|_{\text{Dirac}}
= \frac{1}{4 \pi}\,.
\end{eqnarray}

%================
\section{Shear viscosity for electromagnetic field}
\label{sec_electro}
%================

\subsection{Shear viscosity calculation}

Let us now consider the case of electromagnetic fields in the $ R_{\xi} $ gauge and keep the gauge parameter $ \xi $ arbitrary. The corresponding energy-momentum tensor is well known \cite{Birrell:1982ix}
\begin{eqnarray}\nonumber
T_{\mu\nu} &=& T^{\text{\,M}}_{\mu\nu} + T^{\text{\,G}}_{\mu\nu} + T^{\text{\,ghost}}_{\mu\nu}\,,\\ \nonumber
T^{\text{\,M}}_{\mu\nu} &=&  -F_{\mu\alpha}{F_{\nu}}^{\alpha}+\frac{1}{4} \eta_{\mu\nu} F^2\,, \\ \nonumber
T^{\text{\,G}}_{\mu\nu} &=&  \frac{1}{\zeta} \Big\{ A_{\mu}\partial_{\nu}(\partial A)
+
A_{\nu}\partial_{\mu}(\partial A)-
\eta_{\mu\nu} \big[ A^{\lambda}\partial_{\lambda} (\partial A) +\frac{1}{2}(\partial A)^2 \big] \Big\} \,, \\ 
T^{\text{\,ghost}}_{\mu\nu} &=& \partial_\mu  \bar{c}\partial_\nu c+ \partial_\nu\bar{c}\partial_\mu c 
- \eta_{\mu\nu} \partial_{\rho} \bar{c} \partial^{\rho} c\,,
 \label{EMTvector} \end{eqnarray}
where $ T^{\text{\,M}}_{\mu\nu} $ corresponds to the usual Maxwell energy-momentum tensor, $ T^{\text{\,G}}_{\mu\nu} $ corresponds the gauge-fixing terms, and $ T^{\text{\,ghost}}_{\mu\nu} $ is the energy-momentum tensor of Faddeev-Popov ghosts. The usual notations $ F^2 = F_{\mu\nu}F^{\mu\nu} $ and $ (\partial A)= \partial_{\rho} A^{\rho} $ are used. Propagators in the form of Wightman functions in the coordinate representation are also well known \cite{Faddeev1991-xw}
%\begin{eqnarray} \nonumber
%G_{\mu\nu}(x)&=&\langle 0| 
%A_{\mu}(x) A_{\nu}(0)
%|0\rangle_{M} =
%i \int \frac{d^4 p}{(2 \pi)^4} e^{-ipx} \left[\frac{g_{\mu\nu}}{p^2} 
%+\frac{(1-\zeta) p_{\mu} p_{\nu}}{p^4} \right]\,,
%   \,, \\
%G(x) &=& \langle 0| 
%c(x) \bar{c}(0)
%|0\rangle_{M} =
%i \int \frac{d^4 p}{(2 \pi)^4} e^{-ipx} \frac{1}{p^2} 
%\,,
% \label{propA} \end{eqnarray}
%where the arrangement of poles, as for scalar and Dirac fields, corresponds to the choice of a propagator in the form of a positive-frequency Wightman function. It is not difficult to go to the coordinate representation using formulas (\ref{A2}), (\ref{A4})
\begin{eqnarray}\nonumber
G_{\mu\nu}(x)&=&\langle 0| 
A_{\mu}(x) A_{\nu}(0)
|0\rangle_{M} =
\frac{1}{8 \pi^2}\left( \frac{(1+\zeta)\eta_{\mu\nu}}{x^2-i\varepsilon x_0}+\frac{2(1-\zeta)x_{\mu}x_{\nu}}{(x^2-i\varepsilon x_0)^2} \right) \,, \\
G(x) &=& \langle 0| 
c(x) \bar{c}(0)
|0\rangle_{M} = 
-\frac{1}{4 \pi^2} \frac{1}{x^2-i\varepsilon x_0}
\,,
 \label{propAx} \end{eqnarray}
where the poles, as for scalar and Dirac fields, also raised upward relative to the axis of real time.

\begin{figure*}[!h]
\begin{minipage}{0.7\textwidth}
  \centerline{\includegraphics[width=1\textwidth]{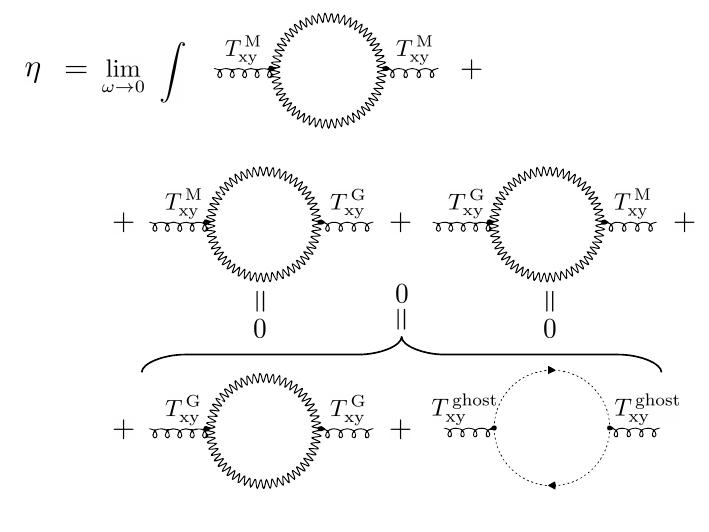}}
\end{minipage}
\caption{Various contributions to photon viscosity, according to expansion (\ref{TTphotonExp}). Contributions equal to zero according to (\ref{TTphotonDifferent}) are shown.}
\label{fig:3}
\end{figure*}

Let's now move on to calculating viscosity. To do this, we first need to calculate the correlator $ \langle \hat{T}\hat{T} \rangle $. Expanding into different contributions from (\ref{EMTvector}), we obtain
\begin{eqnarray} \nonumber
\langle 0| 
\hat{T}_{\mu\nu}(x) \hat{T}_{\alpha\beta}(y)
|0\rangle_{M} &=&
\langle 0| 
\hat{T}^{\text{\,M}}_{\mu\nu}(x) \hat{T}^{\text{\,M}}_{\alpha\beta}(y)
|0\rangle_{M}+
\langle 0| 
\hat{T}^{\text{\,M}}_{\mu\nu}(x) \hat{T}^{\text{\,G}}_{\alpha\beta}(y)
|0\rangle_{M}+
\langle 0| 
\hat{T}^{\text{\,G}}_{\mu\nu}(x) \hat{T}^{\text{\,M}}_{\alpha\beta}(y)
|0\rangle_{M} \\
&&+
\langle 0| 
\hat{T}^{\text{\,G}}_{\mu\nu}(x) \hat{T}^{\text{\,G}}_{\alpha\beta}(y)
|0\rangle_{M} +
\langle 0| 
\hat{T}^{\text{\,ghost}}_{\mu\nu}(x) \hat{T}^{\text{\,ghost}}_{\alpha\beta}(y)
|0\rangle_{M}
\,,
 \label{TTphotonExp} \end{eqnarray}
where we take into account in advance that the off-diagonal terms, where one of the vertices contains $ \hat{T}^{\text{\,ghost}}_{\mu\nu} $ and the other vertex contains the field $ A_{\mu} $, are equal to zero, since we are only interested in connected correlators. The corresponding expansion is shown in the Figure \ref{fig:3}.

In the general case, as already mentioned, for any conformally symmetric theory one should expect that the correlator $ \langle \hat{T}\hat{T} \rangle $ has the form (\ref{I}) up to an overall coefficient. Let's calculate each of the contributions shown in the Figure \ref{fig:3}. To do this, let's present EMT operators in split form\footnote{As before, splitting is not a regularization and only helps to work with derivatives. In this case, there is a certain arbitrariness regarding which field is considered at point $ x_1 $ or $ x_2 $, on which, however, the final answer does not depend.}
\begin{eqnarray} \nonumber 
\hat{T}^{\text{\,M}}_{\mu\nu}(x) &=& \lim_{x_1,x_2\to x} \mathcal{D}^{\text{\,M}}_{\mu\nu\lambda\rho }(\partial_{x_1},\partial_{x_2})A^{\lambda}(x_1)A^{\rho}(x_2)\,, \\
 \nonumber 
\hat{T}^{\text{\,G}}_{\mu\nu}(x) &=& \lim_{x_1,x_2\to x} \mathcal{D}^{\text{\,G}}_{\mu\nu\lambda\rho }(\partial_{x_1},\partial_{x_2})A^{\lambda}(x_1)A^{\rho}(x_2)\,, \\
 \nonumber 
\hat{T}^{\text{\,ghost}}_{\mu\nu}(x) &=& \lim_{x_1,x_2\to x} \mathcal{D}^{\text{\,ghost}}_{\mu\nu}(\partial_{x_1},\partial_{x_2})\bar{c}(x_1)c(x_2)\,, \\
\mathcal{D}^{\text{\,M}}_{\mu\nu\lambda\rho }(\partial_{x_1},\partial_{x_2}) &=& 
\frac{1}{2} \eta_{\mu\nu} (\eta_{\lambda\rho}\partial_{\sigma}^{x_1}\partial_{x_2}^{\sigma} - \partial_{\rho}^{x_1}\partial^{x_2}_{\lambda})+\eta_{\rho\nu}\partial_{\mu}^{x_1}\partial^{x_2}_{\lambda}
-\eta_{\lambda\rho}\partial_{\mu}^{x_1}\partial^{x_2}_{\nu}
-\eta_{\lambda\mu}\eta_{\rho\nu}\partial_{\sigma}^{x_1}\partial_{x_2}^{\sigma}
+\eta_{\lambda\mu}\partial_{\rho}^{x_1}\partial^{x_2}_{\nu}\,,
\nonumber \\
\mathcal{D}^{\text{\,G}}_{\mu\nu\lambda\rho }(\partial_{x_1},\partial_{x_2}) &=&
\frac{1}{\zeta} (\eta_{\mu\lambda}\partial_{\rho}^{x_2}\partial^{x_2}_{\nu}+
\eta_{\rho\nu} \partial_{\mu}^{x_1}\partial^{x_1}_{\lambda}-
\eta_{\mu\nu}\partial_{\lambda}^{x_2}\partial^{x_2}_{\rho} -
\frac{1}{2}\eta_{\mu\nu}\partial_{\lambda}^{x_1}\partial^{x_2}_{\rho})\,, \nonumber
\\
\mathcal{D}^{\text{\,ghost}}_{\mu\nu}(\partial_{x_1},\partial_{x_2}) &=&  \partial_{\mu}^{x_1}\partial_{\nu}^{x_2}+\partial_{\mu}^{x_2}\partial_{\nu}^{x_1}-
\eta_{\mu\nu}\partial_{\sigma}^{x_1}\partial^{\sigma}_{x_2}\,.
\label{TTTDDD}\end{eqnarray}
Wick's theorem takes the form
\begin{eqnarray} \nonumber 
\langle 0 | A_{\mu}(x_1)A_{\nu}(x_2) A_{\alpha}(y_1)A_{\beta}(y_2)|0\rangle_{M,\text{connected}} &=&
\langle 0 | A_{\mu}(x_1)A_{\alpha}(y_1)|0\rangle_{M}  \langle 0 | A_{\nu}(x_2)A_{\beta}(y_2)|0\rangle_{M} \\  \nonumber 
&&+
\langle 0 | A_{\mu}(x_1)A_{\beta}(y_2)|0\rangle_{M}  \langle 0 | A_{\nu}(x_2)A_{\alpha}(y_1)|0\rangle_{M} 
\,, \\
\langle 0 | \bar{c}(x_1)c(x_2) \bar{c}(y_1)c(y_2)|0\rangle_{M,\text{connected}} &=&
\langle 0 |\bar{c}(x_1)c(y_2)|0\rangle_{M}  \langle 0 |c(x_2)\bar{c}(y_1)|0\rangle_{M}\,.
 \label{AAAA}
\end{eqnarray}
Next, we substitute (\ref{TTTDDD}) and (\ref{AAAA}) into (\ref{TTphotonExp}). For each term we obtain
\begin{eqnarray} \nonumber 
\langle 0| 
\hat{T}^{\text{\,M/G}}_{\mu\nu}(x) \hat{T}^{\text{\,M/G}}_{\alpha\beta}(y)
|0\rangle_{M} &=& \lim_{
	\def\arraystretch{0.5}\begin{array}{ll}
	{\scriptscriptstyle x_1,x_2\to x}\vspace{0.1mm}\\ 
	{\scriptscriptstyle y_1,y_2\to y}
	\end{array}} 
\mathcal{D}^{\text{\,M/G}}_{\mu\nu\lambda\rho }(\partial_{x_1},\partial_{x_2})
\mathcal{D}^{\text{\,M/G}}_{\alpha\beta\lambda'\rho' }(\partial_{x_1},\partial_{x_2})
[G_{x_1-y_1}^{\lambda\lambda'}G_{x_2-y_2}^{\rho\rho'} +
G_{x_1-y_2}^{\lambda\rho'} G_{x_2-y_1}^{\rho\lambda'}]\,,  \\ 
\langle 0| 
\hat{T}^{\text{\,ghost}}_{\mu\nu}(x) \hat{T}^{\text{\,ghost}}_{\alpha\beta}(y)
|0\rangle_{M} &=& - \lim_{
	\def\arraystretch{0.5}\begin{array}{ll}
	{\scriptscriptstyle x_1,x_2\to x}\vspace{0.1mm}\\
	{\scriptscriptstyle y_1,y_2\to y}
	\end{array}} 
\mathcal{D}^{\text{\,ghost}}_{\mu\nu}(\partial_{x_1},\partial_{x_2})
\mathcal{D}^{\text{\,ghost}}_{\alpha\beta}(\partial_{x_1},\partial_{x_2}) G_{x_1-y_2}G_{x_2-y_1} 
\,,  \label{TTphotonDD}
\end{eqnarray}
where $ \text{\,M/G} $ means that the vertex is either Maxwellian or the gauge-fixing term (\ref{EMTvector}). Next, we need to substitute the form of $ \mathcal{D} $ operators from (\ref{TTTDDD}) into (\ref{TTphotonDD}), arrange the derivatives, and then go to the limit $ x_1, x_2\to x $ and $ y_1, y_2\to y $. After this we  use the form of the propagators (\ref{propAx}). As a result, we obtain
\begin{eqnarray} \nonumber 
\langle 0| 
\hat{T}^{\text{\,M}}_{\mu\nu}(x) \hat{T}^{\text{\,M}}_{\alpha\beta}(y)
|0\rangle_{M} &=& \frac{16}{\pi^4} \mathcal{I}_{\mu\nu\alpha\beta}(x-y) \,, \\ \nonumber
\langle 0| 
\hat{T}^{\text{\,M}}_{\mu\nu}(x) \hat{T}^{\text{\,G}}_{\alpha\beta}(y)
|0\rangle_{M} &=& 0\,, \\ \nonumber
\langle 0| 
\hat{T}^{\text{\,G}}_{\mu\nu}(x) \hat{T}^{\text{\,M}}_{\alpha\beta}(y)
|0\rangle_{M} &=& 0\,, \\ \nonumber
\langle 0| 
\hat{T}^{\text{\,G}}_{\mu\nu}(x) \hat{T}^{\text{\,G}}_{\alpha\beta}(y)
|0\rangle_{M} &=& \frac{1}{\pi^4}\bigg( \frac{16 b_{\mu} b_{\nu} b_{\alpha} b_{\beta}}{\bar{b}^{12}}
-\frac{2\eta_{\mu\alpha} b_{\nu} b_{\beta}}{\bar{b}^{10}}
-\frac{2\eta_{\nu\alpha} b_{\mu} b_{\beta}}{\bar{b}^{10}}
-\frac{2\eta_{\mu\beta} b_{\nu} b_{\alpha}}{ \bar{b}^{10}}
-\frac{2\eta_{\nu\beta} b_{\mu} b_{\alpha}}{ \bar{b}^{10}}
+\frac{\eta_{\mu\alpha} \eta_{\nu \beta}}{2 \bar{b}^8} 
\\ \nonumber
&& +\frac{\eta_{\mu\beta} \eta_{\nu \alpha}}{2 \bar{b}^8}
+\frac{2\eta_{\mu\nu} \eta_{\alpha \beta}}{\bar{b}^8} -\frac{4\eta_{\mu\nu} b_{\alpha} b_{\beta}}{ \bar{b}^{10}}
-\frac{4\eta_{\alpha\beta} b_{\mu} b_{\nu}}{ \bar{b}^{10}}\bigg) \,,\\ 
\langle 0| 
\hat{T}^{\text{\,ghost}}_{\mu\nu}(x) \hat{T}^{\text{\,ghost}}_{\alpha\beta}(y)
|0\rangle_{M} &=&  - \langle 0| 
\hat{T}^{\text{\,G}}_{\mu\nu}(x) \hat{T}^{\text{\,G}}_{\alpha\beta}(y)
|0\rangle_{M}
\,.  \label{TTphotonDifferent}
\end{eqnarray}
Thus, the mixed contributions with one EMT operator $ \hat{T}^{\text{\,M}}$ and one $ \hat{T}^{\text{\,G}}$ are equal to zero. Also, ghosts and gauge fixing terms mutually cancel each other, as shown in Figure \ref{fig:3}. As a result, only the Maxwellian contribution remains, which satisfies the general formula for the covariant two-point function
\begin{eqnarray}
\langle 0| 
\hat{T}_{\mu\nu}(x) \hat{T}_{\alpha\beta}(y)
|0\rangle_{M} &=&
\langle 0| 
\hat{T}^{\text{\,M}}_{\mu\nu}(x) \hat{T}^{\text{\,M}}_{\alpha\beta}(y)
|0\rangle_{M} = \frac{16}{\pi^4} \mathcal{I}_{\mu\nu\alpha\beta}(x-y)\,.
 \label{TTphoton} \end{eqnarray}

Generally speaking, subsequent calculations need not be performed, since (\ref{TTphoton}) differs from (\ref{TTdirac}) only in the general coefficient and subsequent calculations exactly repeat calculations with the Dirac field (and the scalar field). All that remains is to take into account the difference in the overall coefficient. Thus, we obtain for local viscosity an expression similar to (\ref{etaDiracLoc}) and (\ref{etaScalarLoc})
\begin{eqnarray} \label{etaPhotonLoc}
			\eta^{\text{\,photon}}_{\text{\,loc}} (\rho)=\frac{\rho\left[\rho^4+ 4 \rho^2 l_c^2 -5l_c^4   
				-4 l_c^2(2\rho^2+l_c^2) \ln(\frac{\rho}{l_c})\right]}{20 (\rho^2-l_c^2)^4 \pi^2}\,.
\end{eqnarray}
And the global viscosity, according to (\ref{etaLocDef}), is equal to 
\begin{eqnarray} \label{etaPhoton}
\eta^{\text{\,photon}} = \frac{1}{120 \pi^2 l_c^2}
\,.
\end{eqnarray}
\subsection{Entropy calculation}

Let us now move on to calculating entropy. The energy density and pressure of electromagnetic fields with acceleration and temperature are known \cite{Dowker:1994fi, Frolov:1987dz, Page:1982fm}
\begin{eqnarray} \label{EMTphoton}
\langle \hat{T}^{\text{\,photon}}_{\mu\nu}\rangle = \Big(\frac{\pi^2 T^4}{15}+\frac{T^2 |a|^2}{6}-\frac{11 |a|^4}{240 \pi^2}\Big)\Big(u_{\mu}u_{\nu}-\frac{\Delta_{\mu\nu}}{3}\Big)
\,.
\end{eqnarray}
The corresponding pressure (\ref{pDef}) has the form
\begin{eqnarray} \label{pPhoton}
p = \frac{1}{3}\Big(\frac{\pi^2 T^4}{15}+\frac{T^2 |a|^2}{6}-\frac{11 |a|^4}{240 \pi^2}\Big)
\,.
\end{eqnarray}
We find the entropy density using the formula (\ref{sDef})
\begin{eqnarray} \label{sPhotonTa}
s^{\text{\,photon}}_{\text{\,loc}}(T,|a|)= \frac{4\pi^2 T^3}{45} +\frac{T |a|^2}{9}
\,.
\end{eqnarray}
For the Minkowski vacuum, using (\ref{sDef1}), we obtain
\begin{eqnarray} \label{sPhotonLoc}
s^{\text{\,photon}}_{\text{\,loc}}(T=T_U,|a|=1/\rho)= \frac{1}{15 \pi \rho^3}
\,.
\end{eqnarray}
Entropy per unit area of the horizon, according to (\ref{sDef2}) has the form
\begin{eqnarray} \label{sPhoton}
s^{\text{\,photon}} = \frac{1}{30 \pi l_c^2}
\,.
\end{eqnarray}

\subsection{Shear viscosity to entropy ratio}

The ratio of local viscosity (\ref{etaPhotonLoc}) and entropy (\ref{sDiracLoc}) in the case of photons is described by the same universal function (\ref{etaSloc}), (\ref{f}), as for scalar and Dirac fields
\begin{eqnarray}
\frac{\eta_{\text{\,loc}}}{s_{\text{\,loc}}}(\rho)\Big|_{\text{photon}}= f(\rho/l_c)\,.
\end{eqnarray}
The ratio of total viscosity and entropy satisfies (\ref{main})
\begin{eqnarray} \label{etaSglobPhoton}
\frac{\eta}{s}\Big|_{\text{photon}}
= \frac{1}{4 \pi}\,.
\end{eqnarray}

%================
\section{Bulk viscosity}
\label{sec_bulk}
%================

It is known that the bulk viscosity for a black hole membrane (just the membrane, not the fields above the membrane) is negative \cite{Parikh:1997ma}, which perhaps indicates the inapplicability of the hydrodynamic approximation (valid on large scales) to a black hole, which is a finite size object, for which there is no translation invariance. At the same time, in the case of Rindler space, the area of the horizon is infinite, and it should be expected that in this case there will be no such problem \cite{Chirco:2010xx}. Moreover, assuming that the properties of the membrane correspond to the properties of the fields above the membrane, we do not expect negative bulk viscosity for Unruh radiation too.

A prediction for bulk viscosity (\ref{zs}) was also proposed within the holographic approach. Let us now consider, as before, a quantum fluid above the Rindler horizon. At an arbitrary speed of sound $ c_s $, the bulk viscosity can be found using the following Kubo formula \cite{Jeon:1994if}
\begin{eqnarray}
\zeta&=&\pi \lim_{\omega\rightarrow0} \int_{l_c}^{\infty}\rho\,d\rho\int_{l_c}^{\infty}\rho'\,d\rho'\int_{-\infty}^{\infty}\,dx\,dy\,d\tau e^{i\omega\tau} \langle 0|\hat{\mathcal{P}}(\tau, x, y, \rho)\hat{\mathcal{P}} (0, 0, 0, \rho')|0 \rangle_{\text{M}} \,, \\
\hat{\mathcal{P}} &=& c_s^2 \hat{T}_{0}^{0}   +\frac{1}{3}\hat{T}_{i}^{i}\,.
\label{kuboRindlerBulk0}
\end{eqnarray}
If the speed of sound meets the conformal limit
\begin{eqnarray}
c_s^2=1/3 \quad \Rightarrow \quad \hat{\mathcal{P}} = \hat{T}_{\mu}^{\mu} \,,
\label{csConf}
\end{eqnarray}
then the bulk viscosity can be found from the correlator of two EMT traces
\begin{eqnarray}\label{kuboRindlerBulk1}
\zeta=\frac{\pi}{9} \lim_{\omega\rightarrow0}{\int_{l_c}^{\infty}\rho\,d\rho\int_{l_c}^{\infty}\rho'\,d\rho'\int_{-\infty}^{\infty}\,dx\,dy\,d\tau e^{i\omega\tau}\langle 0|\hat{T}_{\mu}^{\mu}(\tau, x, y, \rho)\hat{T}_{\nu}^{\nu}(0, 0, 0, \rho')|0 \rangle_{\text{M}}} \,.
\end{eqnarray}
For massless Dirac field (\ref{EMTdirac}), electromagnetic (\ref{EMTvector}) and massless scalar field (\ref{EMTscalar}) at $ \xi =1/6 $ the speed of sound satisfies the conformal limit (\ref{csConf}). Indeed, in all these cases the energy density $ \varepsilon= \langle \hat{T}_{\mu\nu} \rangle u^{\mu}u^{\nu} $ and pressure $ p= - \frac{1 }{3}\langle \hat{T}_{\mu\nu} \rangle (g^{\mu\nu}-u^{\mu}u^{\nu}) $ satisfy ultrarelativistic equation of state $ \varepsilon = 3p $, due to which\footnote{For scalar fields at $ \xi \neq 1/6 $ according to (\ref{EMTscalar}), the ultrarelativistic equation of state is violated $ \langle \hat{T}_{\mu}^{\mu}\rangle = \varepsilon - 3 p\neq 0 $. This could be expected in advance, since at $ \xi \neq 1/6 $ the conformal symmetry is broken. Moreover, anisotropy appears - the pressure turns out to be different in different directions: along the acceleration and in the transverse plane. Because of this, generally speaking, one should talk about the existence of two different speeds of sound \cite{Ferrer:2022afu}.}
\begin{eqnarray}\label{сsScDiPh}
c_s^2=\frac{\partial p}{\partial \varepsilon} = \frac{1}{3} \,,
\end{eqnarray}
and we can use the formula (\ref{kuboRindlerBulk1}).

Let us first consider conformally symmetric scalar fields with $ \xi=1/6 $. In this case, the result is trivial and the bulk viscosity vanishes, since already at the operator level (\ref{Tscalar}) the EMT trace is equal to zero
\begin{eqnarray}\label{Zsc}
\text{Scalar:} \quad\hat{T}^{\mu}_{\mu} = 0 \quad \Rightarrow \quad 
\langle 0|\hat{T}_{\mu}^{\mu} \hat{T}_{\nu}^{\nu}|0 \rangle_{\text{M}} =0 \quad \Rightarrow \quad 
\zeta_{\text{\,scalar}} = 0 \,.
\end{eqnarray}

Let us now consider the Dirac fields. The trace of the EMT operator itself is equal to zero only at the classical level on the solutions of the Dirac equation. Therefore, it is necessary to calculate the correlator. Actually, we have already calculated it in (\ref{TTdirac}) and it is expressed through the universal tensor $ \mathcal{I}_{\mu\nu\alpha\beta} $, which has the following obvious properties
\begin{eqnarray}\label{Iprop}
{\mathcal{I}_{\mu}}{}^{\mu}{}_{\alpha\beta} =
\mathcal{I}_{\mu\nu\alpha}{}^{\alpha}= 0
 \,.
\end{eqnarray}
Therefore, we find that the bulk viscosity is zero
\begin{eqnarray}\label{Zd}
\text{Dirac:}\quad \langle 0|\hat{T}_{\mu}^{\mu} \hat{T}_{\nu}^{\nu}|0 \rangle_{\text{M}} =0 \quad \Rightarrow \quad 
\zeta_{\text{\,Dirac}} =0
 \,.
\end{eqnarray}

Since for photons the correlator (\ref{TTphoton}) is the same, up to a general coefficient, as for scalar fields (\ref{TTscalar}) and for Dirac fields (\ref{TTdirac}), it is obvious that due to (\ref{Iprop}) the bulk viscosity will also be zero
\begin{eqnarray}\label{Zph}
\text{Photon:}\quad \langle 0|\hat{T}_{\mu}^{\mu} \hat{T}_{\nu}^{\nu}|0 \rangle_{\text{M}} =0 \quad \Rightarrow \quad 
\zeta_{\text{\,photon}} =0
 \,.
\end{eqnarray}
In this case, already at the operator level, obviously, the Maxwellian part of the EMT immediately drops out, since $ \hat{T}_{\text{\,M,}\mu}^{\mu} =0 $. Thus, compensation occurs for the contributions of the gauge fixing terms and the ghosts
\begin{eqnarray}\label{ZghGauge}
\langle 0|\hat{T}_{\text{\,ghost,}\mu}^{\mu} \hat{T}_{\text{\,ghost,}\nu}^{\nu}|0 \rangle_{\text{M}} = -\langle 0|\hat{T}_{\text{\,G,}\mu}^{\mu} \hat{T}_{\text{\,G,}\nu}^{\nu}|0 \rangle_{\text{M}} = -\frac{12}{\pi^4 \bar{b}^8}
 \,.
\end{eqnarray}
Thus, for fields living above the Rindler membrane, the problem with negative bulk viscosity (which arose for the black hole membrane) has a trivial solution. Let us finally note that according to (\ref{zs}) for $ c_s^2=1/3 $, the lower limit of bulk viscosity is exactly zero, and the bound takes the form $ \zeta \geqslant 0 $. Thus, both shear and bulk viscosity lie simultaneously at the lower limit in all cases considered.

%================
\section{Discussion}
\label{sec_disc}
%================

%================
\subsection{Unruh effect and KSS bound: resume}
\label{subsec_unruh}
%================

%подведение итога: про соединение эффекта Унру и теории струн, про то, что для эффекта Унру это новое, про то, что термальная баня имеет также вязкость
In this work, we considered the relationship between two different well-known effects - the universal KSS-bound (\ref{nsmain}) and the Unruh effect. According to the Unruh effect, a medium similar to Hawking radiation appears in an accelerated frame, which has a finite universal temperature (\ref{TU}). Now we can say that it also has a finite viscosity, the ratio of which to the entropy density is also universal and equal to (\ref{main}). Note that the Unruh effect has been studied in very many works and continues to be the focus of attention (see, for example, \cite{Deur:2024szw}), but mostly only the thermodynamic properties of Unruh radiation are considered. Now we can also say that Unruh radiation is the most ideal fluid with minimal shear viscosity and the problem of detailed study of its hydrodynamic properties arises.

%подчеркивание того, что это прямой расчет, а не голографический
Note that the derivation made did not use the properties of the membrane itself or considerations from holography in any way and is based on the direct application of the concrete quantum field theory and Kubo formulas. Moreover, the derivation \cite{Chirco:2010xx} and ours differs significantly from the well-known membrane derivation \cite{Parikh:1997ma} briefly discussed in Section \ref{sec_kubo}, in which it is essential that viscosity appears as a quantity visible to an infinitely distant observer. And we directly calculate the viscosity at each specific point (see Subsection \ref{subsec_local}) using well-known statistical formulas \cite{zubarev1974nonequilibrium}.

%(про то, что добавка не приводит к "взрыву")
The obtained result raises a number of questions. For example, if we slightly increase the temperature, creating free particles in the inertial frame, then it would seem that this could abruptly lead to infinite viscosity? However, if we approach this problem in this way, then the question arises about the rule for summing viscosities in a two-component-like system, which is quite nontrivial. In particular, if we assume that this rule will be similar to the case with electric current in a parallel circuit, where the inverse resistences are summed, then adding a component with infinite viscosity will not lead to a explosive growth of ``total viscosity''. We hope to explore this issue in more detail in the future.

%================
\subsection{``Entanglement viscosity''}
\label{subsec_entangl}
%================

%про происхождение вязкости - что ожидалось для свободных полей другое
An obvious central question is the origin of the observed dissipative properties. However, the answer is not entirely clear. Indeed, it would seem that we are considering free fields, that is, an ideal gas. Therefore, viscosity has an essentially ``kinematic'' nature, not dynamic (associated with interaction). However, here the question immediately arises - it is well known that the shear viscosity of an ideal gas is formally infinite $ \eta \to \infty $ (see for example \cite{Jeon:1994if}), since it is determined by the mean free path $ \eta \sim l_{\text{\,free}}$, which tends to infinity $ l_{\text{\,free}}\to \infty $. 

However, it is necessary to take into account that we are considering a system in space with a horizon. The existence of a horizon generates entanglement, which, in particular, creates the entanglement entropy. In \cite{Chirco:2010xx}, it was suggested that entanglement is also a source of viscosity (\ref{etaScalar}), (\ref{etaDirac}), (\ref{etaPhoton}), i.e., one can speak about ``entanglement viscosity''.

This view is supported by some indirect indications. In particular, it is known that the horizon entropy can be calculated either using the Bekenstein-Hawking formulas or as the entanglement entropy for fields in a space with a horizon. Although the two methods lead to somewhat similar results (for example, both entropies are proportional to the area of the horizon, see \cite{Ryu:2006bv, Fursaev:2006ih}), the exact relationship between them remains not entirely clear. In particular, while the Bekenstein entropy is universal, the entanglement entropy depends on the type of particles (see, for example, (\ref{sScalar}), (\ref{sDirac}), (\ref{sPhoton})). 

In (\ref{etaSglobScalar}), (\ref{etaSglobDirac}), (\ref{etaSglobPhoton}), actually, just the entanglement entropy appears in the denominator, and we obtain a correspondence with the KSS-bound. It is natural to assume that the value in the numerator is also associated with the entanglement.

Moreover, the ``entanglement viscosity'', like the entanglement entropy, turns out to be different for different types of fields. Indeed, despite the fact that for all the fields considered the ratio of shear viscosity to entropy satisfies (\ref{main}), the viscosities themselves (\ref{etaScalar}), (\ref{etaDirac}), (\ref{etaPhoton}) differ
\begin{eqnarray}
\eta^{\text{\,scalar}}=\frac{1}{6} \eta^{\text{\,Dirac}}= \frac{1}{12} \eta^{\text{\,photon}}=\frac{1}{1440 \pi^2 l_c^2}\,.
\label{nnn}
\end{eqnarray}
In this sense, if viscosities (\ref{nnn}) are similar to entanglement entropy, one can assume that the membrane viscosity calculated on the basis of (\ref{wil}) in \cite{Parikh:1997ma} is more similar, on the contrary, to Bekenstein-Hawking entropy. However, in both cases of ``entanglement viscosity'' and the viscosity of the membrane itself \cite{Parikh:1997ma}, ration (\ref{main}) is fulfilled. So we can say that there is a correspondence or duality between the properties of the membrane and quantum fluid living above this membrane. It would be of interest to study this duality in more detail.

There are other indications of the relationship of the obtained viscosity with the horizon. The existence of a horizon is known to lead to irreversibility - falling into a black hole is irreversible - which is closely related to irreversibility in statistical physics  \cite{Witten:2024upt}. The latter is precisely responsible for dissipation. Thus, apparently, viscosity (\ref{nnn}) is associated with the irreversibility introduced by the Rindler horizon. For example, the existence of a horizon may lead to the situation that one of the particles in the particle-antiparticle pair born from the vacuum may be absorbed by the horizon, forming a fluctuation above the horizon, which could potentially also be a source of frictional forces, as is considered in the context of superfluidity \cite{Landau1980-fa}.

%================
\subsection{Local and global viscosities}
\label{subsec_local}
%================

From our analysis it follows that the relation $ \eta/s =1/4\pi $ is satisfied only for averaged or global quantities (\ref{etaSglobScalar}), (\ref{etaSglobDirac}), (\ref{etaSglobPhoton}). Global viscosity (\ref{etaScalar}), (\ref{etaDirac}), (\ref{etaPhoton}) as well as total entropy (\ref{sScalar}), (\ref{sDirac}), (\ref{sPhoton}), correspond to the unit area of the horizon. Generally speaking, one could expect in advance that the relation (\ref{main}), initially obtained from the physics of black holes, will be associated precisely with the integrated quantities per unit area of the horizon (in \cite{Chirco:2010xx} this statement about the correspondence of the properties of the membrane and physics in volume is introduced as a hypothesis or postulate).

Note that $ \eta \sim 1/l_c^2$, and $ s \sim 1/l_c^2$, diverges as the inverse power of the membrane thickness $ l_c $. Such behavior of entropy $ s\sim 1/l_c^2 $ in (\ref{sScalar}), (\ref{sDirac}), (\ref{sPhoton}) is typical \cite{Solodukhin:2011gn}, and it is well known that, generally speaking, it can depend on the calculation scheme (for example, on how the cutoff of integrals associated with the membrane is introduced). An obvious manifestation of this is the dependence on the $l_c$ parameter. Note, however, that the ratio $ \eta/s $ is apparently an invariant result.
In particular, it does not depend on the parameter $ l_c $, which does not even need to tend to zero and can be left  arbitrary.
\begin{figure*}[!h]
\begin{minipage}{0.7\textwidth}
  \centerline{\includegraphics[width=1\textwidth]{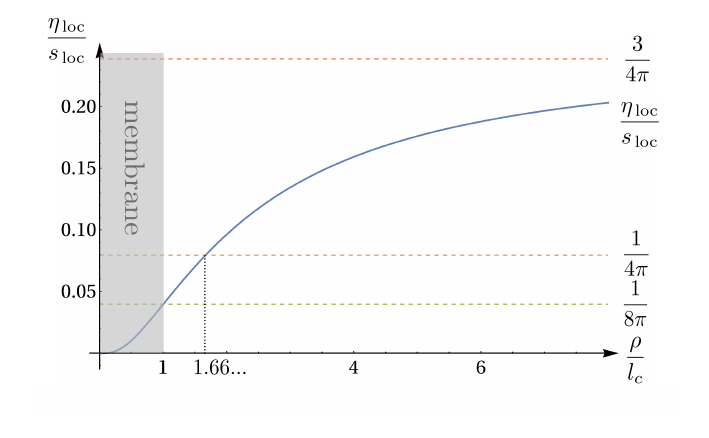}}
\end{minipage}
\caption{The ratio of local viscosity to local entropy according to (\ref{etaSloc}). The corresponding ratio is shown as a blue solid line. Dashed lines indicate concrete values of $ \eta_{\text{\,loc}}/s_{\text{\,loc}} $ at  $ \rho= l_c $ (green), $ \rho=l_0=1.66... \, l_c $ (orange) and $ \rho \to \infty $ (red). The region corresponding to the membrane $ 0<\rho<l_c $ is shaded gray, that is, the area between the extended and true horizons.}
\label{fig:4}
\end{figure*}

However, if we do not take the last integral in (\ref{kuboRindler}) $ \int d\rho $, then the corresponding quantities will describe viscosity and entropy at a certain distance $ \rho $ from the horizon. We called the corresponding quantities ``local shear viscosity'' and ``local entropy''. 

We have written down local viscosities for various fields in the formulas (\ref{etaScalarLoc}), (\ref{etaDiracLoc}), (\ref{etaPhotonLoc}). Similarly, for local entropy the formulas are given in (\ref{sScalarLoc}), (\ref{sDiracLoc}), (\ref{sPhotonLoc}). One can see, that for different fields, viscosity (and entropy) differ only by an overall coefficient, multiplied by the same universal function. The relation of local quantities $ \eta_{\text{\,loc}}/s_{\text{\,loc}} = f(\rho/l_c) $ is described by a rather nontrivial function of the form (\ref{etaSloc}). It is noteworthy that this function turns out to be universal for all types of fields considered (and also, apparently, for a wider class of theories, see Subsection \ref{subsec_gen}).
 
Let us analyze in more detail the dependence of $ \eta_{\text{\,loc}}/s_{\text{\,loc}} $ on the distance to the horizon. The corresponding graph is shown in Figure \ref{fig:4}. Let us list the main conclusions:
\begin{itemize}
\item Locally, the ratio of shear viscosity to entropy can be either greater or less than $ 1/4\pi $. In particular, $ \eta_{\text{\,loc}}/s_{\text{\,loc}} (\rho)<1/4\pi $ for $ \rho<1.66...\, l_c $ (the corresponding value $ l_0= 1.66...\, l_c $ can be found numerically as the root of the equation $ f(l_0/l_c)=1/4\pi $).
\item In particular, on the membrane surface, when $ \rho = l_c $, the ratio is half the KSS bound, that is $ \frac{\eta_{\text{\,loc}}}{s_{\text{\,loc}}} (\rho = l_c ) =\frac{1}{8 \pi} $.
\item  On the contrary, for $ \rho\gg l_c $ the ratio turns out to be higher $ \frac{\eta_{\text{\,loc}}}{s_{\text{\,loc}}} (\rho\to \infty) \to \frac{3}{4\pi}$. Note that this limit does not depend of neither $ \rho $ nor $ l_c $.
\item Analytical continuation inside the membrane shows that at the true horizon the ratio tends to zero $ \frac{\eta_{\text{\,loc}}}{s_{\text{\,loc}}} (\rho\to \infty) \to 0$.
\end{itemize}

In conclusion, we note that the ``local bulk viscosity'', due to (\ref{Zsc}), (\ref{Zd}), (\ref{Zph}), is obviously equal to zero, like the global one, in all the considered cases.
%================
\subsection{Definition of entropy}
\label{subsec_entropy}
%================

There are various approaches to determining entropy in spaces with a horizon \cite{Solodukhin:2011gn, Witten:2024upt}. We followed the ``thermodynamic'' definition from the relativistic spin hydrodynamics  \cite{Becattini:2023ouz, Becattini:2019poj}, according to which in a medium with spin density $ S^{\mu\nu} \equiv u_{\lambda} S^ {\lambda, \mu \nu} $, where $ S^{\lambda, \mu \nu} $ is the mean value of the  spin tensor, the differential thermodynamic relation for pressure is modified
\begin{eqnarray} \label{dp}
dp= s dT + n d\mu + \frac{1}{2} S^{\mu\nu} d\omega_{\mu\nu} \,,
\label{dp_gen}
\end{eqnarray}
where $ \omega_{\mu\nu} $ is the so-called spin potential. This potential, generally speaking, may not be associated with four-velocity gradients; however, in the case we are considering, the system is in global thermodynamic equilibrium \cite{Weert:1982, Becattini:2023ouz}, due to which $ \omega_{\mu \nu} =a_{\mu} u_{\nu} - a_{\nu} u_{\mu} $ (the psedovector component of $ \omega_{\mu \nu} $, vorticity $ \omega_{\mu}=\frac{1}{2} \varepsilon_{\mu\nu\alpha\beta}u_{\nu}\partial^{\alpha} u^{\beta} =0$ equals zero in our case). Assuming also that the chemical potential is zero $ \mu =0$, we obtain from (\ref{dp_gen}) the key formula (\ref{sDef}).

The results we obtained (\ref{etaSglobScalar}), (\ref{etaSglobDirac}), (\ref{etaSglobPhoton}) demonstrate the advantage of this approach to entropy, since it allows us to obtain an exact match with the KSS bound (\ref{main}). 

It is interesting to compare entropy for different types of particles. We obtain, according to (\ref{sScalar}), (\ref{sDirac}), (\ref{sPhoton})
\begin{eqnarray}
s^{\text{\,scalar}}=\frac{1}{6} s^{\text{\,Dirac}}= \frac{1}{12} s^{\text{\,photon}}=\frac{1}{360 \pi l_c^2}\,,
\label{sss}
\end{eqnarray}
which corresponds to the ratios for viscosities (\ref{nnn}).

Note also that there is a significant difference in calculating viscosity and entropy in terms of the methods used. Viscosity can be calculated without going beyond the Minkowski vacuum, as described in sections \ref{sec_scalar}-\ref{sec_electro}, where all results were obtained, in fact, from well-known propagators in the Minkowski vacuum. In contrast to this to calculate entropy according to (\ref{sDef}), we need to deviate a little from the Minkowski vacuum, considering the state with temperature $ T=T_U + dT$ (we only considered the case $ dT>0 $). However, we see that the ``thermal'' calculation of entropy is consistent with the ``vacuum'' calculation of viscosity, leading to the ratio $\eta/s = 1/4\pi$ (however, it is to be expected that problems may arise for higher spins, where the limiting value may differ from the exact one \cite{Fursaev:1996uz}).

%================
\subsection{General case of conformal field theory}
\label{subsec_gen}
%================

Thus, for various field theories - with spins 0, 1/2 and 1 -, the ratio of viscosity to entropy satisfies the KSS bound (\ref{main}). Also, in all the cases considered, the ratio of local quantities (\ref{etaSloc}) turns out to be universal. A natural question is how general is this result?

As was noted earlier, the function
\begin{eqnarray}
 \langle 0| \hat{T}_{\mu\nu}(x) \hat{T}_{\mu\nu}(0) |0 \rangle_{\text{M}} = c \,  \mathcal{I}_{\mu\nu\alpha\beta} (x)\,,
\label{TT_conf}
\end{eqnarray}
has a universal form for any conformal field theory up to a general coefficient $ c $ \cite{Erdmenger:1996yc}. This coefficient is determined by the central conformal charge of the corresponding field theory \cite{Kovtun:2008kw}. 

Considering that, as can be seen from Sections \ref{sec_scalar}-\ref{sec_electro}, all further calculations of viscosity are based only on the use of the expression for the correlator (\ref{TT_conf}), it is obvious that for any conformal theory
\begin{eqnarray}
 \eta \sim c/l_c^2 \,.
\label{eta_conf}
\end{eqnarray}
If the entropy density in the Rindler space were also proportional to the central charge $ s \sim c /l_c^2 $, then one could claim that $ \eta/s=1/4\pi $ holds for any conformal field theory. This is well known in the case of AdS/CFT duality: the entropy of a heated gas is proportional to the conformal charge for conformal theories that have a dual description, based on string or brane theory \cite{Kovtun:2008kw}. However, up to our knowledge, this issue has not been so well studied for the Rindler space.

%Note, however, that the universality of (\ref{main}) and (\ref{etaSloc}) exists not only for conformal theories. It is shown, in Section \ref{sec_scalar} (following \cite{Chirco:2010xx}) that for scalar fields for parameter values $\xi \neq 1/6 $, when conformal symmetry is explicitly broken, we still . 

Finally, since for any conformal field theory the correlator $ \langle 0 | \hat{T}_{\mu\nu} \hat{T}_{\alpha\beta} |0\rangle_{\text{M}} $ has the form (\ref{TT_conf}), for which the convolution is zero (\ref{Iprop}), then the bulk viscosity will also be zero $ \zeta_{\text{\,conf}} = 0 $.

%================
\section{Conclusion \& perspectives}
\label{sec_concl}
%================

There are various views on the essence of the Unruh effect. In particular, it can be considered just as the response of an accelerated (Unruh-DeWitt) detector to the Minkowski vacuum, thus being a property of the detector itself. The results obtained, however, are an additional argument in favor of ``thermal bath view'' that in an accelerated system a real thermal medium appears  (see, for example, \cite{Becattini:2017ljh, Prokhorov:2019cik, Prokhorov:2019yft}), possessing not only a finite temperature, but also nonzero viscosity \footnote{Of course, there is no contradiction between the results of different approaches and the difference appears at the level of interpretations.}. The corresponding viscosity is kinematic in nature and is presumably related to entanglement.

In more detail, we considered free quantum fields living above a membrane describing the horizon of an accelerated reference frame. We have calculated the shear viscosity at the Unruh temperature, i.e. for the Minkowski vacuum, for massless fields with spins 1/2 and 1 (while spin 0 was considered earlier). We have also calculated the entanglement entropy density in all these cases, using the thermodynamic definition through pressure derivative. The ratio of averaged or global quantities, related to a unit area of the horizon, correspond to the lower bound predicted in the KSS-ratio, i.e. $ \eta/s=1/4\pi $. Thus, the universality of this ratio for various field theories is directly shown. Arguments are given in favor of the universality of the relation in the more general case of conformally symmetric theories.

At the same time, the ratio of local quantities characterizing shear viscosity and entropy density at a finite distance $ \rho $ from the Rindler horizon changes as a function of this distance. The horizon is considered as a membrane of finite thickness $ l_c $. In particular, we show, that on the membrane surface the ratio of local shear viscosity to local entropy is less than the KSS bound: $ \eta_{\text{\,loc}}/s_{\text{\,loc}}=1/8\pi $, i.e. half the limit from string theory. And, conversely, asymptotically, at large distances $ \rho\gg l_c $ from the horizon, this ratio is higher than the KSS bound: $ \eta_{\text{\,loc}}/s_{\text{\,loc}}=3/4\pi $. In general, the ratio of local quantities is described by the function, which turns out to be universal for all the considered theories of massless fields (with spins 0, 1/2 and 1).

We also calculated the bulk viscosity, which it turned out to be zero, in contrast to the black hole membrane, where on the ``classical level'' (that is, for the membrane itself) the bulk viscosity was negative.

Thus we see that the limit $ \eta/s=1/4\pi $ is realized precisely for the Minkowski vacuum, e.g. when $ T=T_U $, and for $ T>T_U $, when the state is heated, one could speculate that $ \eta/s > 1/4\pi $, so the limit  $ \eta/s>1/4\pi $ may coincide with the limit $ T>T_U $ for the (minimal) boundary temperature of the accelerated medium \cite{Becattini:2017ljh}. 
%However, taking into account that for the two subsystems it is necessary to sum the inverse viscosities, it the ratio will remain constant. 
On the other hand, the analytic continuation to the region $ T< T_U $ was investigated in \cite{Prokhorov:2023dfg, Prokhorov:2019hif}, where it was shown that $ T=T_U $ is the critical temperature. It can be expected that the KSS-limitation may be violated for $ T< T_U $. 
%However, a direct calculation in this case turns out to be technically more difficult due to the appearance of conical geometry and the presence of a phase transition, which is an essentially non-perturbative effect. 
We plan to explore these possibilities in the future.

Note also that at present, various transport effects arising in moving media studied using quantum-statistical and hydrodynamical methods \cite{Buzzegoli:2017cqy, Sheng:2024pbw, Prokhorov:2018bql, Son:2009tf}. It is shown that they lead to effects observed in heavy-ion collisions  (e.g., in hadronic polarization)%, where extreme acceleration and vorticity can be achieved \cite{Baznat:2017jfj, Drogosz:2024gzv, Palermo:2024tza}
. Recently, in a number of examples \cite{Prokhorov:2022udo, Becattini:2017ljh, Prokhorov:2019yft}, a duality was shown between effects of this kind
% arising in quantum statistical mechanics 
and phenomena in spaces with non-trivial geometry, that is, statistical distributions in Minkowsky space ``know'' about the properties of spaces with horizon and/or curvature. It would be interesting if a similar duality existed for the dissipative properties of a quantum fluid living above a membrane. In particular, the considered case with $ T=T_U $ is expected to be realized in heavy-ion collisions in the boundary region between the core and the corona of a fireball, as was recently shown explicitly by the direct phenomenological modeling \cite{Prokhorov:2025vak}.

%\appendix
%
%%================
%\section{Transition to coordinate representation}
%\label{append}
%%================
%
%In the case of massless fields, it is convenient to switch to the coordinate representation. After that taking loop integrals is reduced to multiplying the propagators and the corresponding vertices. For this one can use the general formula for integrals in dimensional regularization \cite{Grozin:2003ak, Kazakov:2008tr}
%\begin{eqnarray}
%&&\int\frac{d^{D}p\, e^{ipx}}{p^{2(\lambda+1-\alpha)}} = \frac{i 2^{2 \alpha} \pi^{\lambda+1}  \Gamma(\alpha)}{x^{2 \alpha} \Gamma(\lambda+1-\alpha)}\,,\nonumber \\ 
%&&\lambda =1-\varepsilon\,,\quad D=2(\lambda+1)\,.
%\label{A1}
%\end{eqnarray}
%In the cases we are interested in, we obtain
%\begin{eqnarray}\label{A2}
%\int\frac{d^{4}p\, e^{-ipx}}{p^{2}} &=& \frac{4 i \pi^2}{  x^2}\,,\\ 
%\int\frac{d^{4}p\, e^{-ipx} p^{\mu}}{p^{2}} &=& \frac{8 \pi^2 x^{\mu}}{  x^4}\,, \label{A3}\\
%\int\frac{d^{4}p\, e^{-ipx}p^{\mu}p^{\nu}}{p^{4}} &=& \frac{2 i \pi^2}{  x^2}\Big(g^{\mu\nu}-2\frac{x^{\mu}x^{\nu}}{x^2}\Big)\,. 
%\label{A4}
%\end{eqnarray}

%%=================================================
%{\bf Acknowledgements}
%%=================================================
%
%The work was supported by Russian Science Foundation Grant No. 24-22-00124.

\bibliography{lit}

\end{document}